\documentclass[%
reprint,
groupedaddress,
amsmath,amssymb,
aps,pra,
twocolumn,
showpacs,
]{revtex4-1}

\usepackage{tikz}
\usepackage{graphicx}
\usepackage{dcolumn}
\usepackage{bm}
\usepackage{braket}
\usepackage{hyperref}
\usepackage[caption=false]{subfig}
\usepackage{amsmath}
\usepackage{graphicx}
\usepackage{fancyhdr}
\usepackage{float}
\usepackage{siunitx}
\DeclareSIUnit{\belmilliwatt}{Bm}
\DeclareSIUnit{\dBm}{\deci\belmilliwatt}
\DeclareSIUnit{\gammas}{$\Gamma$}
\usepackage{xcolor}
\usepackage{xspace}
\usepackage{eucal}
\usepackage{multirow}
\usepackage[utf8]{inputenc}
\usepackage{pgfplots}
\pgfplotsset{compat=1.14}

\newcommand{\mf}{m_F\xspace}

\newcommand{\DOne}{D\textsubscript{1}\xspace}
\newcommand{\DTwo}{D\textsubscript{2}\xspace}

\newcommand{\Rb}{\textsuperscript{87}Rb\xspace}
\newcommand{\K}{\textsuperscript{39}K\xspace}

\DeclareSIUnit{\gauss}{G}
\begin{document}

\newcommand{\update}[1]{\textbf{\color{red}UPDATE:~#1}\xspace}
\newcommand{\strike}[1]{\color{red}\sout{#1}\xspace}
\newcommand{\new}[1]{{\color{green}#1}\xspace}
\sisetup{detect-weight=true, detect-family=true}

\title{
Rapid generation of all-optical \K Bose-Einstein condensates using a low-field Feshbach resonance
}
\author{A.~Herbst}
\author{H.~Albers}
\author{K.~Stolzenberg}
\author{S.~Bode}
\author{D.~Schlippert}\email[Electronic mail: ]{schlippert@iqo.uni-hannover.de}
\affiliation{Leibniz Universit\"at Hannover, Institut f\"ur Quantenoptik,\\ Welfengarten 1, 30167 Hannover, Germany}

\date{\today}

\begin{abstract}
Ultracold potassium is an interesting candidate for quantum technology applications and fundamental research as it allows controlling intra-atomic interactions via low-field magnetic Feshbach resonances. 
However, the realization of high-flux sources of Bose-Einstein condensates remains challenging due to the necessity of optical trapping to use magnetic fields as free parameter.
We investigate the production of all-optical \K Bose-Einstein condensates with different scattering lengths using a Feshbach resonance near \SI{33}{G}.
By tuning the scattering length in a range between \SI{75}{a_0} and \SI{300}{a_0} we demonstrate a trade off between evaporation speed and final atom number and decrease our evaporation time by a factor of 5 while approximately doubling the evaporation flux.
To this end, we are able to produce fully condensed ensembles with \SI{5.8e4} atoms within \SI{850}{\milli\second} evaporation time at a scattering length of \SI{232}{a_0} and \SI{1.6e5} atoms within \SI{3.9}{\second} at \SI{158}{a_0}, respectively.
We deploy a numerical model to analyze the flux and atom number scaling with respect to scattering length, identify current limitations, and simulate the optimal performance of our setup. 
Based on our findings we describe routes towards high-flux sources of ultra-cold potassium for inertial sensing.
\end{abstract}

\maketitle

\section{Introduction}

Decades after their first experimental realization~\cite{Anderson95Science, Davis95PRL}, Bose-Einstein condensates (BEC) have become a central tool in research ranging from many-body physics~\cite{Bloch2008} to quantum technology applications such as computation~\cite{Ladd2010}, simulation~\cite{Bloch2012}, and sensing and metrology~\cite{Gebbe2021}.
High-flux sources of Bose-Einstein condensates have always been of particular interest, especially with respect to signal-to-noise ratios and quantum projection noise.

To this end, the state-of-the-art has been established by sources based on atom chips~\cite{Rudolph15NJP, Farkas10APB} as demonstrated in the scope of compact apparatuses for microgravity experiments~\cite{Muentinga13PRL,Lachmann21NatComm,Deppner21PRL,Becker2018} and as chosen for current and planned experiments in orbit on the ISS~\cite{Aveline20Nature,Frye2021EPJQT}.
Here, forming traps near the current-conducting structures allows generating strongly confining traps and hence rapid and efficient evaporation.
However, the nearby surface of the atom chip may sometimes be considered unfavorable, e.g., with respect to clipping of atom optics light fields or, in presence of notable temperature gradients, due to black-body radiation~\cite{HaslingerNaturePhys2017}.
In an alternative approach, Bose-Einstein condensation has been demonstrated in all-optical setups~\cite{Barrett01PRL,Clement09PRAR,Landini12PRA,Stellmer13PRA} capable of trapping any magnetic substate.
With the ability to focus optical dipole trap beams into the center of the experimental apparatus, generally this leaves a larger clear aperture for optical access as compared to atom chip solutions~\cite{Kulas2016,Vogt2019}.
Contrary to their magnetic counterparts, the trap depth in optical traps formed by static focused Gaussian beams is inherently linked to the trap's confinement.
Accordingly, lowering the trap depth, as demanded for evaporative cooling, leads to a loss of peak atomic density and elastic scattering rate, thus inhibiting efficient cooling.
Recent studies have shown a variety of tools to counteract these scaling laws~\cite{OHara01PRA}, e.g., by movable lens systems~\cite{Kinoshita05PRAR} enabling a tunable increase in confinement by tighter optical waists or dynamically shaped time-averaged potentials~\cite{Roy2016PRA,Condon2019,Albers2022Commun}.
Finally, while trapping any substate irrespective of whether high or low magnetic field seeking, with the external magnetic field as a free parameter optical traps offer ideal conditions for studying and utilizing Feshbach resonances~\cite{Inouye98Nature,Chin2010}.
As a versatile means of tailoring interactions, Feshbach resonances have for instance enabled the production of cold molecules from a Fermi gas~\cite{Regal2003,Cubizolles03PRL}, molecular BECs~\cite{Jochim03Science,Zwierlein03PRL,Greiner2003}, sympathetic cooling~\cite{Roati2007PRL,Campbell2010PRA}, ground-state molecules~\cite{Voges20PRL}, and studies of interaction dynamics~\cite{Zhang2021,Eigen2018}.
Finally, for cooling fermionic species, Feshbach-induced collisions have been proposed and demonstrated as a evaporation knife~\cite{Mathey09PRA,Peng21arXiv}.

In this work, we demonstrate rapid evaporation of \K gases to quantum degeneracy from a time-averaged optical trap. 
By using low-field Feshbach resonances~\cite{DErrico07NJP,Landini12PRA,Salomon14PRA} we are able to directly tune the rethermalization rate and show a dependence between evaporation flux in the BEC and the scattering length when sufficiently distant from the resonance.
Accordingly, compared to the largest prepared BEC we are able to decrease our evaporation time by a factor of 5 while approximately doubling the evaporation flux when tuning the scattering length.
By generating Feshbach magnetic fields via readily existing coils employed for our magneto-optical trap, our approach might be implemented more easily in other existing setups compared to previously demonstrated hybrid trap techniques~\cite{Landini12PRA}.
We describe our apparatus and the detailed experimental sequence including direct loading of the optical trap and a multi-stage magnetic state preparation.
We finally discuss the prospects of using our result in compact quantum inertial sensors using tunable interactions.

\section{Experimental setup}

\begin{figure*}[t!]
    \begin{center}
    \includegraphics[width=1\textwidth]{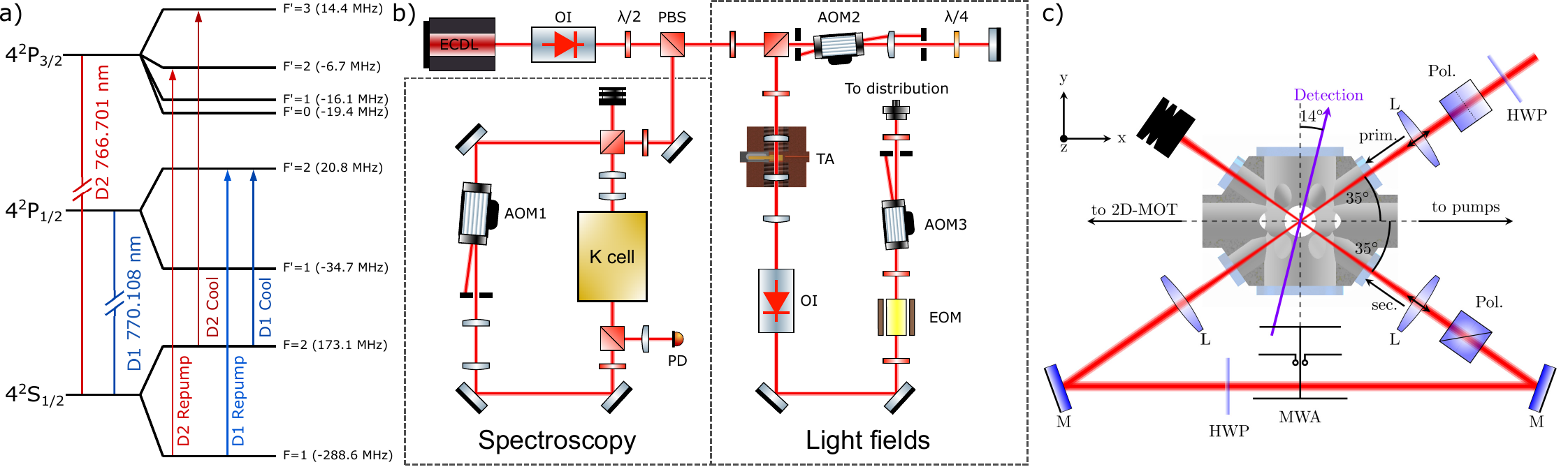}
    \caption{
    \textbf{a) Level structure of $^{39}$K:} For trapping and cooling of potassium, transitions on both the \DOne-line and the \DTwo-line are used. 
    \textbf{b) Layout of the laser system for \DOne-cooling:} The external cavity laser (ECDL) is stabilized via modulation transfer spectroscopy to the $\ket{F=1/2}\rightarrow\ket{F'=1/2}$ crossover transition. The double-pass AOM (AOM2) shifts the laser's frequency to match the \DOne cooling transition before passing the amplifier (TA). Optical isolators (OI) are used to prevent potential back-reflections into TA and ECDL. The free space EOM creates the repump light. The subsequent AOM (AOM3) allows for fast amplitude switching.
    \textbf{c) Optical dipole trap layout:} The primary laser beam (prim.) is focused, re-collimated, redirected and focused a second time (sec.) through the vacuum chamber by three lenses (L) with a focal length of \SI{150}{\milli\metre}.
    To suppress heating effects the linear polarization of the laser is \SI{90}{\degree} rotated  by $\lambda/2$ retardation plates (HWP) and cleaned up by two laser polarizers (Pol.), which are oriented orthogonal to each other.
    A Yagi-Uda micro-wave antenna (MWA) is used for coherent manipulation of the atoms. Quantization coil pairs are aligned with the depicted coordinate system, while the Feshbach fields are generated by another Helmholtz pair along the y-axis.
    The detection (purple arrow) is performed under an angle of \ang{14} in respect to the y-axis.
    }
    \label{fig:Laser_system}
    \end{center}
\end{figure*}

\subsection{Laser system}

Trapping and cooling of potassium is performed using transitions on both the \DOne-line at \SI{770.108}{\nano\meter} and the \DTwo-line at \SI{766.701}{\nano\meter} (Fig.~\ref{fig:Laser_system}a). 
The \DTwo-laser system and its performance have been described previously~\cite{Schlippert14PRL, Albers2020EPJD}.
In this work we additionally operate a laser system on the \DOne-line (Fig.~\ref{fig:Laser_system}b).
To this end, \SI{5}{\milli\watt} of an external cavity diode laser's~\cite{Baillard06OC,Gilowski07OC} output are used for frequency stabilization to the potassium \DOne-crossover line $\ket{F=1/2}\rightarrow\ket{F'=1/2}$ in a vapour cell by means of modulation transfer spectroscopy~\cite{McCarron2008}.
The remaining \SI{30}{\milli\watt} are frequency shifted by a double-pass AOM to match the \DOne cooling transition $\ket{F=2}\rightarrow\ket{F'=2}$ which has an offset of \SI{152.3}{\mega\hertz} with respect to the crossover transition.
Subsequently, the light is amplified by a tapered amplifier [Eagleyard EYP-TPA-0765-01500-3006-CMT03-0000] yielding a total output power of \SI{1.5}{\watt}.
The repump transition $\ket{F=1}\rightarrow\ket{F'=1}$ is addressed using a side band generated by a resonant free-space EOM [Qubig PM-K39+41] driven with a maximum input power of \SI{31.7}{\dBm}, allowing one to freely tune the power ratio between carrier and side band in a range from $0$ to $1$.
The cooling and repumping fields are then transported via a polarization-maintaining fiber [S\&K PMC-E-630-4.1-NA012-3-APC.EC-1000-P] with \SI{60}{\%} coupling efficiency yielding \SI{500}{\milli\watt} in total at the fiber output. 
Finally, the fiber output is superimposed with the \DTwo-light using a narrow-line interference filter and distributed between MOT and detection fibers.

\subsection{Optical dipole trap}

Our optical dipole trap is based on a \SI{1960}{\nano\meter} fiber laser [IPG, TLR-50-1960-LP] operated at \SI{38}{\watt} output power.
The lasers intensity is stabilized by a FPGA feedback loop controlling and linearizing a Pockels cell and analyzer setup, which reduces the power by approximately \SI{40}{\percent}.
Subsequent the light passes an acousto-optical modulator (AOM) [Polytec ATM-1002FA53.24, custom made] with \SI{60}{\percent} diffraction efficiency.

We then focus the elliptical beam in the vacuum chamber with a beam waist of 30 (45)~\si{\micro\meter} in horizontal (vertical) direction and realize a recycled cross under an angle of \SI{70}{\degree}~(Fig.~\ref{fig:Laser_system}c) with a maximum power of \SI{8}{\watt} in the primary and \SI{6}{\watt} in the secondary beam.
The high power losses are mainly caused by imperfect optical elements available for the wavelength of \SI{1960}{\nano\meter}.

The previously mentioned AOM is used to modulate the center-position of the laser beam.
The modulation reaches amplitudes of up to \SI{200}{\micro\meter} in the primary and \SI{300}{\micro\meter} in the secondary recycled beam.
By this we generate time-averaged potentials~\cite{Roy2016PRA,Albers2022Commun,Albers2020phd} in the horizontal plane.
The shape of the resulting potential depends on the modulation of the AOM driving radio frequency, which is generated using a voltage-controlled oscillator [Mini-Circuits, ZOS-150+] controlled by the output of an arbitrary waveform generator [Rigol, DG1022Z].
The waveform of the control voltage is chosen to generate preferably parabolic potentials with a modulation frequency of \SI{20}{\kilo\hertz}.
The amplitude of the waveform defines the spatial width of the center-position modulation of the laser beam and is controlled using a modulation input on the arbitrary waveform generator.

With this we can reach trap depths from $U_0=$~\SI{130}{\nano\kelvin} to \SI{530}{\micro\kelvin}, corresponding to trapping frequencies of $\omega/2\pi=\{4; 6; 50\}$~\si{\hertz} to $\{1.3; 1.9; 2.2\}$~\si{\kilo\hertz} in $\{x; y; z\}$-direction.

\subsection{Magnetic field control}

The coil setup creating the magnetic fields consists of three Helmholtz pairs oriented along the coordinate system in figure \ref{fig:Laser_system}b) and the main coil pair along the y-axis.
The former are used to define the quantization axes during state preparation and imaging as well as for compensating stray magnetic fields ($\left|B_{x,y,z}\right|\leq$\SI{1}{\gauss}) during cooling and trapping. The stabilization electronics allow for two different operation modes. In presence of the MOT- and Feshbach magnetic fields the system is stabilizing the current, utilizing current transducers [LEM CASR 15-NP]. 
During the state preparation the magnetic fields are directly stabilized using a three-axis fluxgate sensor [Bartington  MAG-03IE1000] with a measurement range of $\pm\SI{10}{\gauss}$ mounted close to the vacuum chamber giving us a  short-term stability of $\sigma_{\{x,\,y,\,z\}}(\tau)= \left\{3.9;\, 2.2;\, 1.4\right\}\,\cdot 10^{-5}\,$\SI{}{\gauss} at $\tau=$ \SI{1}{\second} and $B_{x,\,y,\,z}=$ \SI{1}{\gauss}.

The main coils can be operated in Helmholtz and anti-Helmholtz configuration providing the homogeneous magnetic field for tuning the interaction strength during evaporation and the gradient magnetic fields~\cite{Zaiser2010PHD} in the laser cooling stages. 
To switch between these configurations a home build H-bridge, based on mosfets [IXYS 747-IXFN170N65X2] with a low drain-source on-resistance of \SI{13}{\milli\ohm} is used.
The coils are driven in series by one power supply [EA-PSI 9200-25] with a maximum voltage of \SI{200}{\volt} limiting the current to \SI{11.1}{\ampere} due to the coils internal resistance (\SI{18}{\ohm}). For current stabilization the power supply's internal stabilization loop is used stating a current stability of \SI{<0.15}{\%}.  

For characterizing the applied magnetic field at the position of the atoms we perform an initial estimation using a Biot-Savart model of our setup. 
Subsequently, we determine the coil current needed for several Feshbach resonances by observing the related enhancement of atom losses~\cite{Weber03PRL,Fedichev1996PRL}. 
By comparing the corresponding field estimates to the literature values of the resonances we identified the six resonances expected \cite{DErrico07NJP,Tiemann20PRR} (Table~\ref{tab:Resonances}) in our setup for a mixture of $\ket{F=1,\mf=-1}$ and $\ket{F=1,\mf=0}$ at \SI{15}{\micro\kelvin} below \SI{200}{\gauss}. 
Ultimately, the coil currents and magnetic field literature values are used for the final calibration yielding \SI{41.2}{\gauss/ \ampere} with a maximum magnetic field strength of \SI{450}{\gauss} and an accuracy of \SI{0.2}{\gauss/ \ampere}. 

\begin{table}[h!]
\centering
\caption{Identified Feshbach resonances, used to calibrate the magnetic fields. The magnetic field values are taken from Ref.~\cite{Tiemann20PRR}.}
\begin{tabular}{c c c c}
\hline \hline
 & &  &  Literature \\ 
\multicolumn{3}{l}{Atom pair ($F,\mf$)}& magnetic field (G)   \\ \hline
$(1,-1)$&$+$& $ (1,-1)$   & $32.6 \pm 1.5$  \\
$(1,0) $&$+$& $ (1,0)$   & $59.3 \pm 0.6$  \\
$(1,0) $&$+$& $ (1,0)$   & $66.0 \pm 0.9$  \\
$(1,0) $&$+$& $ (1,-1)$  & $113.76 \pm 0.1$  \\
$(1,-1)$&$+$& $ (1,-1)$   & $162.8 \pm 0.9$ \\
\hline \hline

\end{tabular}
\label{tab:Resonances}
\end{table}

\section{Quantum gas production}

\subsection{Loading sequence}

\begin{figure*}[t!]
    \begin{center}
    \includegraphics[width=1\textwidth]{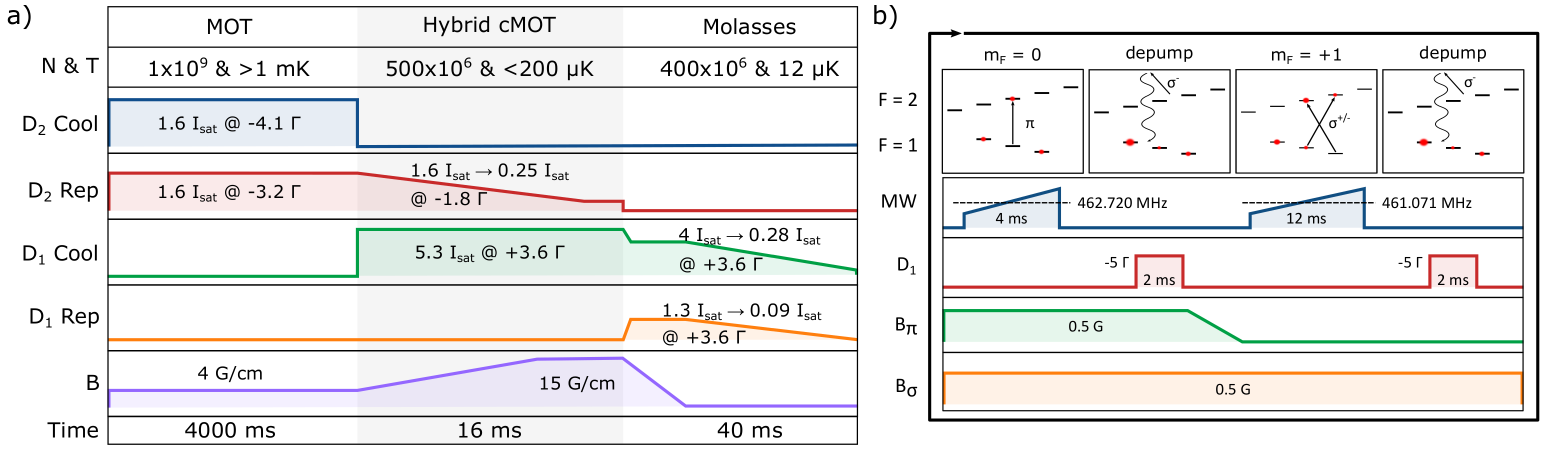}
    \caption{
    \textbf{a) Loading scheme of the optical dipole trap:}  We initially load a MOT on the \DTwo-line from a 2D-MOT. Afterwards we use a Hybrid \DOne -\DTwo compressed MOT to increase the peak density while cooling the ensemble simultaneously. Subsequently sub-Doppler temperatures of \SI{12}{\mu\kelvin} are achieved by \DOne gray molasses cooling. All stated intensities refer to one beam, respectively. This scheme allows to directly load the ODT, without the need for magnetic trapping as intermediate step.
    \textbf{b) State preparation loop within the optical dipole trap:}
    A combination of microwave adiabatic rapid passages and optical pumping is used to shift the population towards $\mf=-1$, starting from a equally populated mixture in $F=1$.  A blow-away sequence is added at the end to increase the purity of the ensemble. After 4 loops, each \SI{70}{\milli\second} long, we reach an almost pure ensemble ($>$98\%) with a temperature increase of \SI{~0.3}{\micro\kelvin} per loop, as determined via time-of-flight measurements. 
    }
    \label{fig:Sequence}
    \end{center}
\end{figure*}

Our dipole trap loading scheme~(Fig.~\Ref{fig:Sequence}a) comprises three steps.
Initially, we load \num{1e9} atoms in a 3D magneto-optical trap (MOT) within \SI{4}{\second} from a beam formed by a 2D-MOT.
Afterwards we switch off the \DTwo cooling light and simultaneously turn on the \DOne cooling light field, thus enabling a \DOne gray molasses dipole trap loading scheme, which is particularly robust against AC stark shifts, as induced by the optical dipole trap~\cite{Salomon13EPL}: 
By ramping up the magnetic field gradient from \SI{4}{\gauss/cm} to \SI{15}{\gauss/cm} and the \DTwo repumping light down to 0.25 I$_{\text{sat}}$ we perform a hybrid \DOne-\DTwo compressed MOT, to increase the density and simultaneously cool the ensemble to below \SI{200}{\micro\kelvin} within \SI{16}{\milli\second}. 
We then switch off the remaining \DTwo repumping light and the magnetic field gradient to perform gray molasses cooling with a \DOne cooling/repumping ratio of 3:1, ramping the cooling (repumping) intensity per beam from 4 (1.3) I$_{\text{sat}}$ to 0.28 (0.09) I$_{\text{sat}}$. 
The ramp length is optimized experimentally to take the decay time of the magnetic gradient field into account. 
The coldest ensembles are realized for a total ramp length of \SI{40}{\milli\second}.
In free space, i.e. in absence of the optical dipole trap, our cooling sequence yields a final temperature of \SI{7}{\micro\kelvin}. 

For loading the optical dipole trap we maximize the number of atoms trapped by adjusting the beam balancing and magnetic offset fields, thus optimizing the overlap at the expense of the final temperature which then yields \SI{12}{\micro\kelvin}.
We find the optimal loading parameters of the optical dipole trap by trading off trap depth $U_0$ and trap volume.
Assuming an initial cloud of $N_0$ atoms with Gaussian density distribution $D_G(x)$ of width $\sigma$ and energy distribution
\begin{equation}
    W(E) = \frac{1}{\pi E k_B T}e^{-\frac{E}{k_B T}},
\end{equation}
the ration of atoms loaded can be estimated for the 1-dimensional case by \cite{Gaaloul2006PRA} 
\begin{equation}
    p_N=\frac{1}{N_0}\int_0^{U_0}\int_{-s_0}^{s_0}D_G(x)W(E)\, dx\, dE,
\end{equation}
with the spatial limits $s_0=\sqrt{\frac{\omega^2}{2}\ln(\frac{U_0}{E})}$.
Primarily, $p_N$ depends on the ratio of trap depth to temperature $U_0/k_B T$ and the width of the trap compared to the atomic ensemble $\omega/\sigma$.
Thus, at constant optical power increasing the center-position modulation (CPM) amplitude and accordingly the width $\omega$ increases the number of atoms trapped, until the trap becomes too shallow and the number of atoms decreases.
With the optical dipole trap already turned on at the maximum available power during the MOT loading and cooling sequence we load a maximum of \num{12e6} atoms into the trap at a CPM amplitude of \SI{160}{\micro\meter}. 
This corresponds to a transfer efficiency of \SI{2}{\percent}, with \num{3e6} atoms positioned in the beams' crossing region.
At this stage we are limited by the trap's depth of \SI{54}{\micro\kelvin} resulting in a final temperature of the trapped atoms of \SI{8.5}{\micro\kelvin}, determined via a time-of-flight measurement after \SI{100}{\milli\second} of rethermalization time. 

\subsection{State preparation}

Upon loading the dipole trap we depump the atomic ensemble into $\ket{F=1}$ by shining in \SI{1}{\milli\watt} of \DOne cooling light with a detuning of $\SI{-1}{\Gamma}$ via the detection optics, thus creating an equidistributed $\mf$-mixture. 
To generate a spin-polarized ensemble in $\ket{F=1, \mf=-1}$, we adapt a multi-loop state preparation scheme~\cite{Antoni-Micollier2017} combining optical pumping on the \DOne-line with coherent transfers~(Fig.~\Ref{fig:Sequence}b). 
While the original publication makes use of optical Raman pulses, we utilize microwave adiabatic rapid passages driven by a Yagi-Uda-type directed antenna.
In order to use a minimal number of loops in the final sequence we experimentally optimize each step towards the highest transfer efficiency.

We initially apply a quantization field of \SI{0.7}{\gauss} at an angle of \SI{45}{\degree} to the antenna's Poynting vector oriented in the horizontal plane.
This allows us to transfer the population of $\ket{F=1,\mf=0}$ to $\ket{F=2,\mf=0}$ by driving a $\pi$-transition sweeping the microwave from \SI{461.710}{\mega\hertz} to \SI{461.730}{\mega\hertz} within \SI{4}{\milli\second}. 
We subsequently apply $\sigma^{-}$-polarized \DOne cooling light via the detection optics with a detuning of $\SI{-5}{\Gamma}$ for \SI{2}{\milli\second} in order to depump the atoms into $\ket{F=1}$ via spontaneous emission, hence populating $\ket{F=1, \mf=0}$ and $\ket{F=1, \mf=-1}$.  
Due to the mismatch of quantization field axis and detection beam orientation by \SI{14}{\degree} we also drive the $\ket{F=2, \mf=-2}\rightarrow\ket{F'=2, \mf=-2}$ transition, thus preventing accumulation in $\ket{F=2, \mf=-2}$.
We then rotate the magnetic field axis to be parallel to the antenna's Poynting vector by tuning the vertical field component to zero, effectively resulting in an quantization field in the horizontal plane of \SI{0.5}{\gauss}.
In this configuration we are able to drive a $\sigma$-polarized microwave adiabatic rapid passage.
Sweeping from \SI{462.062}{\mega\hertz} to \SI{462.080}{\mega\hertz} in \SI{12}{\milli\second} we transfer atoms from $\ket{F=1, \mf=+1}$ to $\ket{F=2, \mf=0}$. Due to the symmetry of the level structure the same microwave pulse also drives the unwanted transition from $\ket{F=1, \mf=0}$ to $\ket{F=2, \mf=-1}$. However, since $\ket{F=1, \mf=0}$ has already been addressed with the first pulse fewer atoms are available for undesired transitions. 
We afterwards apply the same optical depumping pulse as before, thus transferring the atoms back into $\ket{F=1}$. 
While this sequence does not yield a pure magnetic substate, it shifts the $\mf$-distribution towards $\mf=-1$.
Repeating the sequence multiple times allows us to accumulate atoms in $\ket{F=1, \mf=-1}$.
As a final step, we purify the ensemble by a blowaway sequence:
We apply the same microwave pulses as before but tune the \DOne cooling laser closer to resonance ($\SI{-1}{\Gamma}$) to purposely heat atoms out of the trap during optical pumping. 

We find the necessary number of loops by optimizing towards the highest atom number in the BEC for a given evaporation ramp, observing no beneficial effects for more than 4 loops as described. 
In this configuration we reach an almost pure ensemble (\SI{>98}{\%} in $\ket{F=1,\mf=-1}$) with \SI{70}{\%} of the initial atoms remaining in the trap.
To determine the heating from optical pumping we perform a time-of-flight measurement out of the ODT before the state preparation sequence and after each loop.
Since the \DOne-line does not feature closed transitions, with \SI{~0.3}{\micro\kelvin} per loop the observed heating is minimal and yields a final temperature of \SI{9.6}{\micro\kelvin}. 

\subsection{Feshbach resonances \& evaporation}

To address Feshbach resonances we generate homogeneous magnetic fields by switching the MOT magnetic field coils from anti-Helmholtz to Helmholtz configuration. We then ramp up the magnetic field to the desired value in \SI{100}{\milli\second}.
The corresponding scattering length close to the resonances can be calculated using the background scattering length $a_{\text{bg}}$, the resonance widths $\Delta_i$ and its center $B_{0i}$ \cite{Tiesinga1993PRA, DErrico07NJP}:
\begin{equation}
    a(B)= a_{\text{bg}}\left(1-\sum_i\frac{\Delta_i}{B-B_{0i}}\right)
    \label{eqn:Feshbach}
\end{equation}

Once a stable final magnetic field and thus atomic scattering length is established we use the dipole trap's intensity stabilization in combination with the AOM to piecewise decrease the trap power in five linear ramps.
In parallel, the center-position modulation is ramped down and reaches zero with the final ramp at the end of the evaporation sequence.
The initial center-position modulation amplitude is determined by the optimal mode match for trap loading. 
We optimize each step with respect to the final atom number in the BEC by tuning the intensity reduction, the CPM amplitude reduction, the ramp lengths as free parameters.

\section{Experimental results}

\begin{figure*}[t!]
    \begin{center}
    \includegraphics[width=1\textwidth]{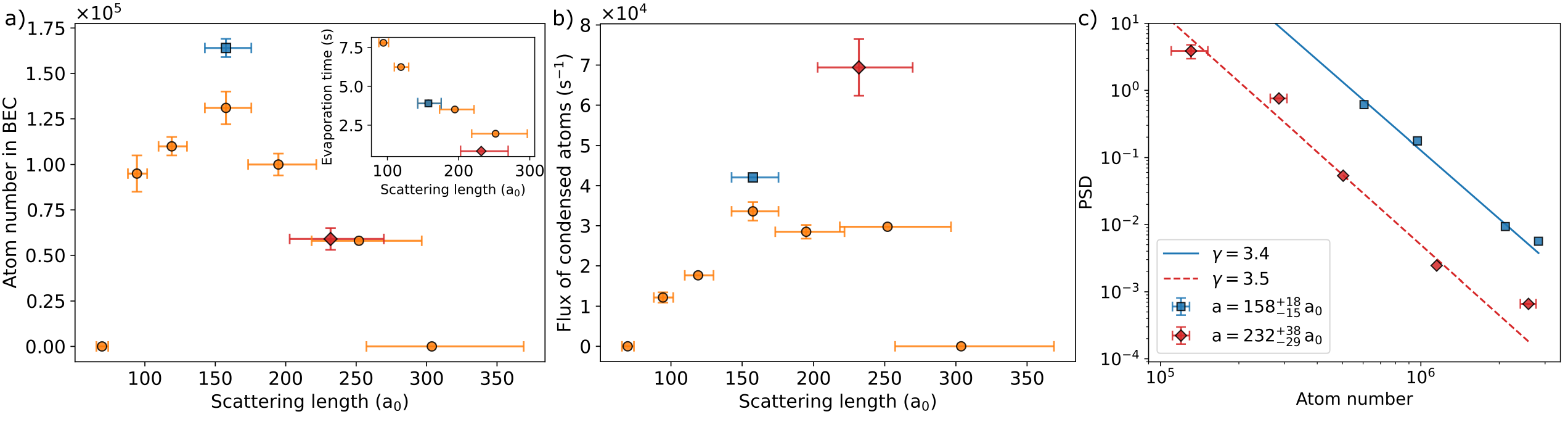}
    \caption{
    \textbf{a) BEC size and time dependency with respect to scattering length:}
    The orange circles show the results, when a given evaporation ramp is re-scaled towards the highest atom number in the BEC at a given scattering length. 
    For a free optimization, including all variables described in the Section III C, the largest fully condensed BEC is realized at a scattering length of \SI{158}{a_0} (blue square). 
    For lower and higher scattering lengths the final atom number is reduced. The inlay shows the time needed for evaporating. For higher scattering lengths the higher elastic collision rates allow for faster evaporation resulting in a smaller time constant. 
    The fastest BEC is realized at a scattering length of \SI{232}{a_0} (red diamond) with a total evaporation time of \SI{850}{\milli\second}. 
    The error bars for the scattering length are determined via error propagation using equation \ref{eqn:Feshbach} and the magnetic field accuracy stated in Section II C. 
    \textbf{b) Flux of the evaporative cooling against scattering length:} 
    The highest flux is achieved for the fastest evaporation ramp at a scattering length of \SI{232}{a_0}. The error bars for the flux are derived via error propagation using the uncertainty in atom number, resulting in larger values for shorter evaporation times.  
    \textbf{c) Phase space density trajectory for largest BEC and highest flux:} 
    For the evaporation efficiency $\gamma$ only the data points on the ramp and not the starting point are taken into account. 
    We find equal efficiency for both cases but deteriorated initial conditions at \SI{232}{a_0}, due to the magnetic field spending an extended time at the resonance resulting in additional heating.} 
    \label{fig:Results}
    \end{center}
\end{figure*}

We systematically optimize BEC production for different scattering lengths~(Fig.~\ref{fig:Results}a). Initially, we optimize the ramp itself.
In accordance with Ref.~\cite{Roy2016PRA} we find our evaporation sequence to be most efficient using an exponential reduction for the trap power jointly with a linear reduction slope for the center-position modulation amplitude, as this allows to counter-act the reduction in trap frequency from the power reduction.

We achieve a fully condensed BEC of \num{1.3e5} atoms at a scattering length of \SI{158}{a_0} after evaporating over the course of \SI{3.9}{\second}.
Contrary to findings in Ref.~\cite{Landini12PRA}, where dynamical tuning, i.e., starting at a lower scattering length and subsequently increasing it, counteracted light-assisted collisions caused by a multimode fiber laser at \SI{1070}{\nano\meter}, we do not observe similar beneficial effects with respect to the final atom number in the BEC. This behaviour qualitatively agrees with experiments demonstrating the absence of light-induced collisions at larger trap wavelengths, e.g., \SI{1550}{\nano\meter} in Ref.~\cite{Salomon14PRA}.
To analyze the scaling behavior with respect to scattering length, we additionally optimize the BEC size at lower and higher scattering length, by re-scaling the overall length of evaporation ramp while maintaining its shape (Fig.~\ref{fig:Results}, orange circles).
At a lower scattering length the time needed for evaporation is increased, since the evaporation rate $\Gamma_{\text{ev}}$ scales with the scattering length squared \cite{Ketterle96AAMOP}
\begin{equation}
    \Gamma_{\text{ev}} = 4\pi a^2 n_0\bar{\nu}\eta e^{-\eta},
    \label{eqn:gamma_ev}
\end{equation}
thus stretching the length of the overall sequence.
Additionally, the final atom number is reduced due to the trapped atoms' lifetime of \SI{15}{\second}.
At scattering lengths \SI{\le 75}{a_0} this effect becomes dominant and we are unable to reach condensation.
For scattering lengths \SI{>158}{a_0} we also find a reduced final atom number, which we explain by the enhancement of the three-body loss rate $\Gamma_{\text{3b}}$, scaling with the scattering length to the power of four \cite{Weber03PRL}:
\begin{equation}
    \Gamma_{\text{3b}}= \frac{K_3}{N}\int n^3d^3r\quad\text{with}\quad K_3 = n_l C(a)\frac{\hbar}{m}a^4.
   \label{eqn:gamma_3b}
\end{equation}
However, the higher evaporation rate now allows one to significantly reduce the time needed for evaporation. 
We experimentally find a limit for this trade off at \SI{>300}{a_0} where the three-body losses become dominant and inhibit Bose-Einstein condensation.
Based on the initial findings we individually alter the relative length of the five linear ramps towards the globally largest BEC size (Fig.~\ref{fig:Results}, blue square) and the largest BEC size at the shortest evaporation time possible (Fig.~\ref{fig:Results}, red diamond).
Realizing the best ratio of desired to undesired losses, we find a maximum atom number of \num{1.6e5} in the BEC at \SI{158}{a_0}, close to the initial configuration and with the same total ramp length.
Our fastest evaporation is performed within \SI{850}{\milli\second} at a scattering length of \SI{232}{a_0} yielding a fully condensed BEC of \num{5.8e4} atoms. 
Compared to the trajectory at \SI{158}{a_0} this corresponds to a reduction in evaporation time by a factor of five while approximately doubling the evaporation flux, which we define as final atom number in the condensate over evaporation time ~(Fig.~\ref{fig:Results}b). 

For the phase space density $\rho$, measured after each linear ramp, we find the evaporation efficiency $\gamma = -\frac{d\ln(\rho)}{d\ln(N)}$  
to be the same for both cases due to the individual optimization (Fig.~\ref{fig:Results}c), but with worse initial conditions at \SI{232}{a_0}.

\section{Discussion}

\subsection{Current limitations}

Currently, our experiment is limited by three effects: Thermal lensing in the optical dipole trap, the stability of the Feshbach fields and worse initial conditions for higher scattering lengths. 
Thermal lensing is a common problem for setups operating laser beams at higher power. 
The effects of all optical elements involved have been discussed, concluding that the dominant contribution originates from the TeO$_2$-crystal used in AOMs \cite{Simonelli2019OptExpress}.
Operating a laser at \SI{1960}{\nano\meter} wavelength, the problem is further amplified, as most of our optics need to be custom-made and typically are subject to higher absorption compared to components for standard wavelengths, e.g. at \SI{1560}{\nano\meter} or \SI{1064}{\nano\meter} (see Section II B).  
To quantify the effect, we displace the secondary dipole trap beam, realizing two parallel, tilt-free single beam traps. 
Measuring the center position of the trapped atoms at different points of the evaporation sequence, we observe a focus drift of up to \SI{200}{\micro\meter} (\SI{450}{\micro\meter}) for the primary (secondary) beam. 
From the experimental optimization we find the ramps at higher scattering length and shorter time scales to be more sensitive regarding deviations from the optimal evaporation trajectory. 
Thus, the reduction in control of the trap parameters from thermal lensing, together with the magnetic field instability, as outlined in Section II C, imposes an upper limit on the scattering length that can be used effectively in our setup. 

Additionally, we observe deteriorating initial conditions when operating closer to the resonance, thus limiting the final atom number in the BEC.
When ramping to the desired magnetic field value after loading the ODT, we sweep across the resonance and hence induce a loss of atoms and heating.
For higher scattering lengths the required magnetic field values are positioned closer to the resonance and the resulting slower sweeps in its direct vicinity amplify the deteriorating effects described above. 
In the explicit case of the trajectory at \SI{232}{a_0} for the 1-$\sigma$ band of our magnetic field estimate there is a five times longer overlap with the resonance compared to the trajectory at \SI{158}{a_0}.
As a result, the initial phase space density is reduced by an order of magnitude.

\subsection{Performance comparison}

State-of-the-art atomic sources generate an evaporation flux of $\ge\SI{2e5}{\per\second}$ with total experimental cycle times at the order of $\sim$\SI{1}{\second} for atom interferometry applications utilizing \Rb~\cite{Rudolph15NJP}. 
Using \K a comparable evaporation flux of \SI{2.4e5}{\per\second} has been demonstrated in a hybrid setup with a magnetic trap as intermediate step, requiring a magnetic field gradient of \SI{270}{\gauss/\centi\meter} and an experimental cycle time of \SI{15}{\second} due to magnetic transport~\cite{Landini12PRA}.   
All-optical cooling allows for shorter cycle times and a simplified apparatus with respect to the generation of magnetic fields.
In our setup, the current density necessary for generating \SI{1}{G/\centi\meter} gradient in anti-Helmholtz configuration generates $\sim$\SI{20}{\gauss} when switched to Helmholtz configuration, thus putting the low-field Feshbach resonances in \K readily in reach with standard quadrupole field coils as used for laser cooling.
For a similar all-optical setup an experimental cycle time of \SI{7}{\second} has previously been reported and yielded an evaporation flux of $\sim$\SI{1e4}{\per\second} with a final atom number of \num{2e4} in the pure condensate~\cite{Salomon14PRA}.
At \SI{232}{a_0} (\SI{158}{a_0}) we realize an evaporation flux of \SI{6.8e4}{\per\second}(\SI{4.1e4}{\per\second}) improving on this result by a factor of $\sim$\num{7} ($\sim$\num{4}).
Accordingly, our final atom number of \num{5.8e4} (\num{1.6e5}) corresponds to an improvement by a factor of $\sim$\num{3} ($\sim$\num{8}). 
To this end our repetition rate is limited by the MOT loading and the data transfer after detection, resulting in a comparable cycle time of \SI{6}{\second} (\SI{9}{\second}). 

\section{Outlook}

\subsection{Further enhancement of atomic flux}

To achieve results comparable to \Rb chip traps, the experimental cycle time and the evaporative flux need further enhancement.  
By using more sophisticated loading techniques such as a high-flux 2D$^+$-MOT design \cite{Catani2006PRA,Landini12PRA} or even cryogenic buffer-gas beam sources~\cite{Doyle2021PRA}, we expect to reduce the MOT loading time to below \SI{1}{\second}, theoretically enabling cycle times of $\sim$\SI{2}{\second}.
Similarly, an upgraded ODT setup would allow to trap more atoms by using a higher beam power together with a larger center-position modulation amplitude, extending the crossing region over the whole molasses without further reducing the trap depth. 
Additionally, by altering the trapping beam's path such that the gray molasses can be used in free space configuration the temperature during loading could be improved by a factor of two.
Regarding the deteriorated phase space density caused by the sweep time of the magnetic fields, possible mitigation strategies include the use of dedicated fast control loops or initially sweeping to a higher field value, allowing to cross the Feshbach resonance with the same sweep rate irrespective of the final scattering length.  

\begin{figure*}[t!]
\begin{minipage}[c]{0.48\textwidth}
    \begin{center}
    \includegraphics[width=1\textwidth]{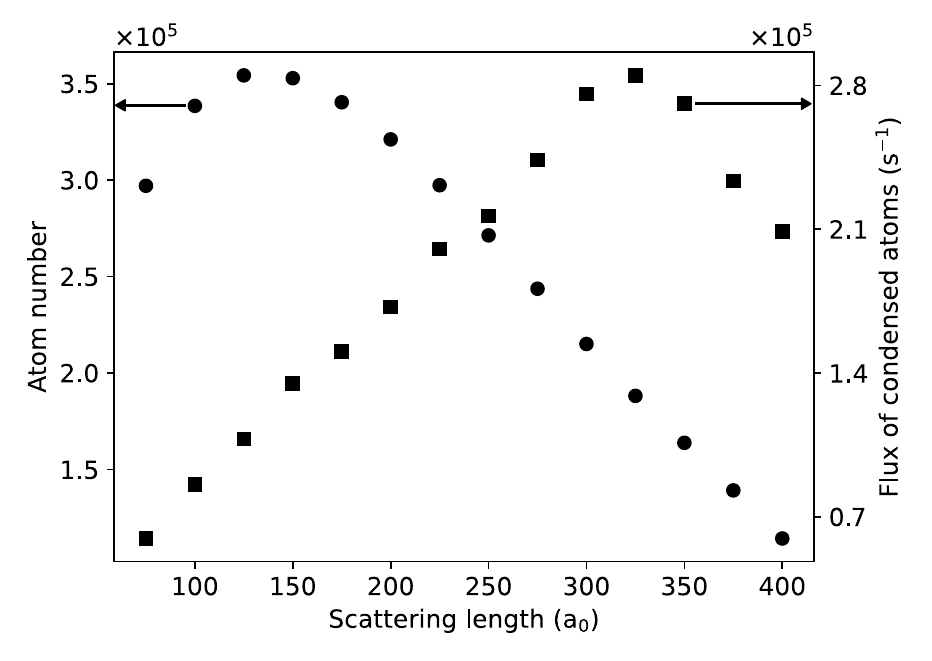}
    \caption{\textbf{Extrapolated performance, simulated for ideal experimental conditions:} 
    For each data point 2000 evaporation trajectories are evaluated. Similar to the experimental optimization, the results for the trajectory yielding the highest final atom number are depicted. Qualitatively, we find the same behaviour as observed in the experiment with improved numbers.   
    } 
    \label{fig:Simulations}
    \end{center}
\end{minipage}
\hfill
\begin{minipage}[c]{0.48\textwidth}
    \begin{center}
    \vspace{4.pt}
    \includegraphics[width=0.97\textwidth]{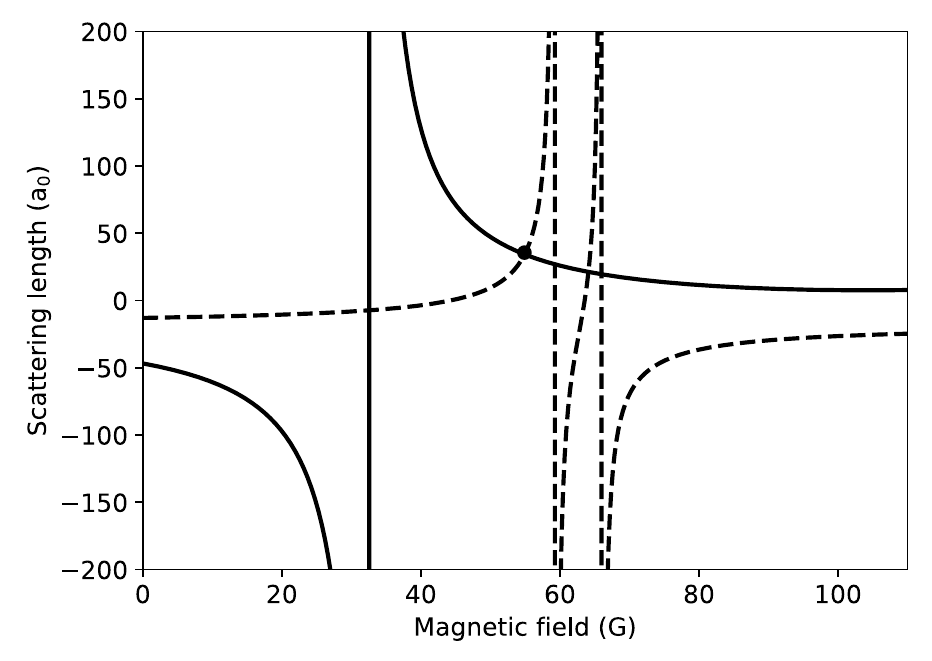}
    \caption{\textbf{Modelling of the scattering length with respect to magnetic field:} Low-field Feshbach resonances with parameters from \cite{Tiemann20PRR} for $\ket{1,-1}+\ket{1,-1}$ (solid line) and $\ket{1,0}+\ket{1,0}$ (dashed line), according to equation~(\ref{eqn:Feshbach}). At \SI{54.9}{\gauss} both resonances share the same scattering length (dot), allowing to easily transfer a BEC via radio frequency adiabatic rapid passages between both states.} 
    \label{fig:Feshbach_resonances_0_1}
    \end{center}
\end{minipage}
\end{figure*}

To explore the theoretical limits of our existing setup we perform additional simulations without the current limitations as identified in Section V A.
We follow the model described in Ref.~\cite{Roy2016PRA}, incorporating equation \ref{eqn:gamma_ev} and \ref{eqn:gamma_3b}.
The optical potential is generated by two identical beams with \SI{8}{W} initial power, a vertical waist of \SI{50}{\micro\meter}, a horizontal waist of \SI{28}{\micro\meter} and ideally overlapping foci.
We assume \num{3e6} atoms to be trapped in the crossing region initially with a temperature of \SI{10}{\micro\kelvin}. 
Evaporation trajectories are defined by an exponential power and linear CPM amplitude reduction and simulated for a given scattering length on a two dimensional grid given by evaporation time and final beam power.
The simulation is aborted once a BEC fraction \SI{>90}{\percent} is estimated using 
\begin{equation}
    \frac{N_c}{N} = 1 -\left(\frac{T}{T_c}\right)^3
\end{equation}
and final atom number and evaporation time are adjusted accordingly. 
Similar to the experiment we analyze the results with the highest evaporation efficiency (Fig.~\ref{fig:Simulations}). 
The simulation results in a peak atom number of \num{3.5e5} at \SI{125}{a_0} after \SI{3.5}{\second} of evaporative cooling. 
Towards higher scattering lengths the final atom number declines due to the reduced ratio of desired to undesired losses, necessitating additional evaporative cooling.
A peak flux of \SI{2.3e5}{\per\second} is simulated at \SI{325}{a_0}, allowing for a final atom number of \num{1.5e5} after \SI{640}{\milli\second}.
Under optimal conditions we thus expect the extrapolated flux to be competitive with results achieved in hybrid traps~\cite{Landini12PRA}.
While resolving each of the identified limitations comes with its own technological challenges, our findings indicate  the potential of the discussed techniques to realize a \K optical dipole trap on par with the current performance of \Rb chip traps.   
 
Since the increase of evaporation flux in the BEC only depends on the accessibility of sufficiently broad Feshbach resonances this scheme could also be utilized for different elements. 
Potentially useful Feshbach resonances are found for \Rb (at $\sim\SI{1007}{\gauss}$ with $\Delta=\SI{0.17}{\gauss}$) \cite{Marte2002PRL}, $^{85}$Rb (at $\sim\SI{155}{\gauss}$ with $\Delta=\SI{11.6}{\gauss}$) \cite{Roberts1998PRL} and $^{23}$Na (at $\sim\SI{90}{\gauss}$ with $\Delta=\SI{1}{\gauss}$) \cite{Knoop2011PRA}, but are subject to individual challenges regarding their width and absolute field magnitude.

\subsection{Application to atom interferometry} 

To apply our methods to atom interferometry a magnetically insensitive population in $\mf=0$ is desired, requiring a state transfer either before or after evaporation.  
If the atoms are already prepared in $\mf=0$, evaporation can be performed in the vicinity of the low-field resonance at \SI{59.3}{\gauss} (Fig.~\ref{fig:Feshbach_resonances_0_1}).
A suitable multi-loop preparation scheme has been demonstrated in Ref.~\cite{Antoni-Micollier2017} and yields comparable results regarding the final purity of the ensemble and the number of required loops.
So far, direct evaporative cooling in $\mf=0$ has been shown on the broad resonance at \SI{471}{\gauss}, with a final atom number in the BEC comparable to the results achieved in $\mf=-1$ ~\cite{Landini12PRA}.
However, due to the narrower shape of the resonance at \SI{59.3}{\gauss} a magnetic field of \SI{58.56}{\gauss} with an instability below \SI{20}{\milli\gauss} is needed in order to achieve same scattering length stability as in our present experiment.
This corresponds to an improvement by a factor of \num{5}, potentially requiring substantial changes to the coil and control setup.  
Moreover, the associated three-body loss coefficient has experimentally been found to be an order of magnitude larger than the one belonging to $\mf=-1$ at \SI{32.6}{\gauss}, thus rendering efficient evaporation unlikely \cite{Semeghini2018PRL,Cabrera2018Science}.

Alternatively, the atoms can be transferred to $\mf=0$ via a radio frequency adiabatic rapid passage after producing the BEC in $\mf=-1$, a technique, which is commonly used for producing quantum droplets \cite{Semeghini2018PRL,Cabrera2018Science}.
Starting from the magnetic field values used in our experiment, a direct transfer would result in an atomic interaction quench to a lower value (Fig. \ref{fig:Feshbach_resonances_0_1}), leading to unwanted density fluctuations \cite{Hung2013Science}.
Instead, for upcoming experiments we plan to sweep the magnetic field to \SI{54.9}{\gauss} where both $\mf$-states share the same scattering length, allowing to transfer the atoms with a radio frequency sweep at $\sim$\SI{40}{\mega\hertz}.

We envisage applications of our source in atom interferometry for inertial sensing.
On one hand, supporting high-fidelity beam splitting and exquisite control of systematic effects high-flux BEC sources are of direct interest in free falling light-pulse atom interferometry~\cite{Hensel2021,Schlippert2020}.
In addition, BECs offer routes to create entanglement via one-axis twisting dynamics~\cite{Corgier21PRA} or delta-kick squeezing techniques~\cite{Corgier21PRL} and routes to transfer spin squeezing to momentum states have been demonstrated~\cite{Anders21PRL}. 
For the specific case of \K, one approach to circumvent the BEC's inherent instability in absence of Feshbach magnetic fields during the interferometer is to exploit ballistic expansion with subsequent matter-wave lensing~\cite{Albers2022Commun,Deppner21PRL,Kovachy15PRL} to operate at low densities.
On the other hand, applications in trapped interferometry using optical guiding potentials will benefit from tunable interactions~\cite{Fattori08PRL,Kim22arXiv}, e.g., to mitigate phase diffusion due to collisions.
Compared to free falling interferometry, here an additional challenge is imposed by the necessity to operate at a constant Feshbach field.
To this end, even with the low-field resonances present in \K, special care needs to be taken to avoid magnetic field gradients and curvature, e.g. by careful design and the use of trimming coils.
Likewise, the use of magnetic shields may pose challenges regarding their compatibility with regular switching of large magnetic fields in their proximity but could be mitigated by active magnetic field control over the comparably small volume of interest of a few cubic millimeters.
Recently demonstrated results from a guided \K multi-loop atom interferometer with interrogation time on the order of milliseconds in presence of an additional axial magnetic potential with trapping frequencies of \SI{2.8}{\hertz} originating from the Helmholtz coils operated near the resonance at \SI{562}{\gauss} \cite{Kim22arXiv} pose a promising step towards using resonances at lower field values where the field curvature can be reduced and an enhanced interrogation time can be expected.

\section{Conclusion}

We have demonstrated an all-optical high flux source of \K BECs. By altering the scattering length via a low-field Feshbach resonance, the evaporation sequence can be individually optimized either towards atom number in the condensate or evaporation flux. In combination with using time-averaged optical potentials out technique allows to improve on results previously achieved with all-optical setups either in atom number by a factor of 8 or evaporation flux by a factor of 7. Therefore, our source's performance is comparable to other high flux sources using \Rb \cite{Kinoshita05PRAR} or $^{174}$Yb \cite{Roy2016PRA}.
We envisage applications in the field of atom interferometry, where measurements are ultimately bound by quantum projection noise. Here, its Feshbach resonances position \K in an interesting spot regarding the possibilities to mitigate systematic effects and dephasing due to collisional shifts at comparably low fields.

\begin{acknowledgments}
We are indebted to Ernst Rasel for inspiring discussions on the general scope of the project and 
Torben Schulze for fruitful input on the use of potassium Feshbach resonances in our experiment.
We thank Carsten Klempt for discussions on the application of our method to other species and Torben Schulze, Ludwig Mathey, Christian Schubert, and Naceur Gaaloul for constructive comments on our manuscript.
This work is funded by the Federal Ministry of Education and Research (BMBF) through the funding program Photonics Research Germany under contract number 13N14875.
The authors further acknowledge support by the German Space Agency~(DLR) with funds provided by the Federal Ministry for Economic Affairs and Energy~(BMWi) due to an enactment of the German Bundestag under Grant No. DLR 50WM2041~(PRIMUS-IV) and by the Deutsche Forschungsgemeinschaft (DFG, German Research Foundation)–Project-ID 274200144–the SFB 1227 DQ-mat within Project No. B07 and under Germany’s Excellence Strategy—EXC-2123 QuantumFrontiers—Project-ID 390837967.
\end{acknowledgments}

\bibliography{main}

\begin{thebibliography}{74}%
\makeatletter
\providecommand \@ifxundefined [1]{%
 \@ifx{#1\undefined}
}%
\providecommand \@ifnum [1]{%
 \ifnum #1\expandafter \@firstoftwo
 \else \expandafter \@secondoftwo
 \fi
}%
\providecommand \@ifx [1]{%
 \ifx #1\expandafter \@firstoftwo
 \else \expandafter \@secondoftwo
 \fi
}%
\providecommand \natexlab [1]{#1}%
\providecommand \enquote  [1]{``#1''}%
\providecommand \bibnamefont  [1]{#1}%
\providecommand \bibfnamefont [1]{#1}%
\providecommand \citenamefont [1]{#1}%
\providecommand \href@noop [0]{\@secondoftwo}%
\providecommand \href [0]{\begingroup \@sanitize@url \@href}%
\providecommand \@href[1]{\@@startlink{#1}\@@href}%
\providecommand \@@href[1]{\endgroup#1\@@endlink}%
\providecommand \@sanitize@url [0]{\catcode `\\12\catcode `\$12\catcode
  `\&12\catcode `\#12\catcode `\^12\catcode `\_12\catcode `\%12\relax}%
\providecommand \@@startlink[1]{}%
\providecommand \@@endlink[0]{}%
\providecommand \url  [0]{\begingroup\@sanitize@url \@url }%
\providecommand \@url [1]{\endgroup\@href {#1}{\urlprefix }}%
\providecommand \urlprefix  [0]{URL }%
\providecommand \Eprint [0]{\href }%
\providecommand \doibase [0]{http://dx.doi.org/}%
\providecommand \selectlanguage [0]{\@gobble}%
\providecommand \bibinfo  [0]{\@secondoftwo}%
\providecommand \bibfield  [0]{\@secondoftwo}%
\providecommand \translation [1]{[#1]}%
\providecommand \BibitemOpen [0]{}%
\providecommand \bibitemStop [0]{}%
\providecommand \bibitemNoStop [0]{.\EOS\space}%
\providecommand \EOS [0]{\spacefactor3000\relax}%
\providecommand \BibitemShut  [1]{\csname bibitem#1\endcsname}%
\let\auto@bib@innerbib\@empty
\bibitem [{\citenamefont {Anderson}\ \emph {et~al.}(1995)\citenamefont
  {Anderson}, \citenamefont {Ensher}, \citenamefont {Matthews}, \citenamefont
  {Wieman},\ and\ \citenamefont {Cornell}}]{Anderson95Science}%
  \BibitemOpen
  \bibfield  {author} {\bibinfo {author} {\bibfnamefont {M.}~\bibnamefont
  {Anderson}}, \bibinfo {author} {\bibfnamefont {J.}~\bibnamefont {Ensher}},
  \bibinfo {author} {\bibfnamefont {M.}~\bibnamefont {Matthews}}, \bibinfo
  {author} {\bibfnamefont {C.}~\bibnamefont {Wieman}}, \ and\ \bibinfo {author}
  {\bibfnamefont {E.}~\bibnamefont {Cornell}},\ }\href {\doibase
  10.1126/science.269.5221.198} {\bibfield  {journal} {\bibinfo  {journal}
  {Science}\ }\textbf {\bibinfo {volume} {269}},\ \bibinfo {pages} {198}
  (\bibinfo {year} {1995})}\BibitemShut {NoStop}%
\bibitem [{\citenamefont {Davis}\ \emph {et~al.}(1995)\citenamefont {Davis},
  \citenamefont {Mewes}, \citenamefont {Joffe}, \citenamefont {Andrews},\ and\
  \citenamefont {Ketterle}}]{Davis95PRL}%
  \BibitemOpen
  \bibfield  {author} {\bibinfo {author} {\bibfnamefont {K.~B.}\ \bibnamefont
  {Davis}}, \bibinfo {author} {\bibfnamefont {M.-O.}\ \bibnamefont {Mewes}},
  \bibinfo {author} {\bibfnamefont {M.~A.}\ \bibnamefont {Joffe}}, \bibinfo
  {author} {\bibfnamefont {M.~R.}\ \bibnamefont {Andrews}}, \ and\ \bibinfo
  {author} {\bibfnamefont {W.}~\bibnamefont {Ketterle}},\ }\href {\doibase
  10.1103/physrevlett.74.5202} {\bibfield  {journal} {\bibinfo  {journal}
  {Physical Review Letters}\ }\textbf {\bibinfo {volume} {74}},\ \bibinfo
  {pages} {5202} (\bibinfo {year} {1995})}\BibitemShut {NoStop}%
\bibitem [{\citenamefont {Bloch}\ \emph {et~al.}(2008)\citenamefont {Bloch},
  \citenamefont {Dalibard},\ and\ \citenamefont {Zwerger}}]{Bloch2008}%
  \BibitemOpen
  \bibfield  {author} {\bibinfo {author} {\bibfnamefont {I.}~\bibnamefont
  {Bloch}}, \bibinfo {author} {\bibfnamefont {J.}~\bibnamefont {Dalibard}}, \
  and\ \bibinfo {author} {\bibfnamefont {W.}~\bibnamefont {Zwerger}},\ }\href
  {\doibase 10.1103/revmodphys.80.885} {\bibfield  {journal} {\bibinfo
  {journal} {Reviews of Modern Physics}\ }\textbf {\bibinfo {volume} {80}},\
  \bibinfo {pages} {885} (\bibinfo {year} {2008})}\BibitemShut {NoStop}%
\bibitem [{\citenamefont {Ladd}\ \emph {et~al.}(2010)\citenamefont {Ladd},
  \citenamefont {Jelezko}, \citenamefont {Laflamme}, \citenamefont {Nakamura},
  \citenamefont {Monroe},\ and\ \citenamefont {O'Brien}}]{Ladd2010}%
  \BibitemOpen
  \bibfield  {author} {\bibinfo {author} {\bibfnamefont {T.~D.}\ \bibnamefont
  {Ladd}}, \bibinfo {author} {\bibfnamefont {F.}~\bibnamefont {Jelezko}},
  \bibinfo {author} {\bibfnamefont {R.}~\bibnamefont {Laflamme}}, \bibinfo
  {author} {\bibfnamefont {Y.}~\bibnamefont {Nakamura}}, \bibinfo {author}
  {\bibfnamefont {C.}~\bibnamefont {Monroe}}, \ and\ \bibinfo {author}
  {\bibfnamefont {J.~L.}\ \bibnamefont {O'Brien}},\ }\href {\doibase
  10.1038/nature08812} {\bibfield  {journal} {\bibinfo  {journal} {Nature}\
  }\textbf {\bibinfo {volume} {464}},\ \bibinfo {pages} {45} (\bibinfo {year}
  {2010})}\BibitemShut {NoStop}%
\bibitem [{\citenamefont {Bloch}\ \emph {et~al.}(2012)\citenamefont {Bloch},
  \citenamefont {Dalibard},\ and\ \citenamefont
  {Nascimb{\`{e}}ne}}]{Bloch2012}%
  \BibitemOpen
  \bibfield  {author} {\bibinfo {author} {\bibfnamefont {I.}~\bibnamefont
  {Bloch}}, \bibinfo {author} {\bibfnamefont {J.}~\bibnamefont {Dalibard}}, \
  and\ \bibinfo {author} {\bibfnamefont {S.}~\bibnamefont {Nascimb{\`{e}}ne}},\
  }\href {\doibase 10.1038/nphys2259} {\bibfield  {journal} {\bibinfo
  {journal} {Nature Physics}\ }\textbf {\bibinfo {volume} {8}},\ \bibinfo
  {pages} {267} (\bibinfo {year} {2012})}\BibitemShut {NoStop}%
\bibitem [{\citenamefont {Gebbe}\ \emph {et~al.}(2021)\citenamefont {Gebbe},
  \citenamefont {Siem{\ss}}, \citenamefont {Gersemann}, \citenamefont
  {Müntinga}, \citenamefont {Herrmann}, \citenamefont {Lämmerzahl},
  \citenamefont {Ahlers}, \citenamefont {Gaaloul}, \citenamefont {Schubert},
  \citenamefont {Hammerer}, \citenamefont {Abend},\ and\ \citenamefont
  {Rasel}}]{Gebbe2021}%
  \BibitemOpen
  \bibfield  {author} {\bibinfo {author} {\bibfnamefont {M.}~\bibnamefont
  {Gebbe}}, \bibinfo {author} {\bibfnamefont {J.-N.}\ \bibnamefont
  {Siem{\ss}}}, \bibinfo {author} {\bibfnamefont {M.}~\bibnamefont
  {Gersemann}}, \bibinfo {author} {\bibfnamefont {H.}~\bibnamefont
  {Müntinga}}, \bibinfo {author} {\bibfnamefont {S.}~\bibnamefont {Herrmann}},
  \bibinfo {author} {\bibfnamefont {C.}~\bibnamefont {Lämmerzahl}}, \bibinfo
  {author} {\bibfnamefont {H.}~\bibnamefont {Ahlers}}, \bibinfo {author}
  {\bibfnamefont {N.}~\bibnamefont {Gaaloul}}, \bibinfo {author} {\bibfnamefont
  {C.}~\bibnamefont {Schubert}}, \bibinfo {author} {\bibfnamefont
  {K.}~\bibnamefont {Hammerer}}, \bibinfo {author} {\bibfnamefont
  {S.}~\bibnamefont {Abend}}, \ and\ \bibinfo {author} {\bibfnamefont {E.~M.}\
  \bibnamefont {Rasel}},\ }\href {\doibase 10.1038/s41467-021-22823-8}
  {\bibfield  {journal} {\bibinfo  {journal} {Nature Communications}\ }\textbf
  {\bibinfo {volume} {12}} (\bibinfo {year} {2021}),\
  10.1038/s41467-021-22823-8}\BibitemShut {NoStop}%
\bibitem [{\citenamefont {Rudolph}\ \emph {et~al.}(2015)\citenamefont
  {Rudolph}, \citenamefont {Herr}, \citenamefont {Grzeschik}, \citenamefont
  {Sternke}, \citenamefont {Grote}, \citenamefont {Popp}, \citenamefont
  {Becker}, \citenamefont {Müntinga}, \citenamefont {Ahlers}, \citenamefont
  {Peters}, \citenamefont {Lämmerzahl}, \citenamefont {Sengstock},
  \citenamefont {Gaaloul}, \citenamefont {Ertmer},\ and\ \citenamefont
  {Rasel}}]{Rudolph15NJP}%
  \BibitemOpen
  \bibfield  {author} {\bibinfo {author} {\bibfnamefont {J.}~\bibnamefont
  {Rudolph}}, \bibinfo {author} {\bibfnamefont {W.}~\bibnamefont {Herr}},
  \bibinfo {author} {\bibfnamefont {C.}~\bibnamefont {Grzeschik}}, \bibinfo
  {author} {\bibfnamefont {T.}~\bibnamefont {Sternke}}, \bibinfo {author}
  {\bibfnamefont {A.}~\bibnamefont {Grote}}, \bibinfo {author} {\bibfnamefont
  {M.}~\bibnamefont {Popp}}, \bibinfo {author} {\bibfnamefont {D.}~\bibnamefont
  {Becker}}, \bibinfo {author} {\bibfnamefont {H.}~\bibnamefont {Müntinga}},
  \bibinfo {author} {\bibfnamefont {H.}~\bibnamefont {Ahlers}}, \bibinfo
  {author} {\bibfnamefont {A.}~\bibnamefont {Peters}}, \bibinfo {author}
  {\bibfnamefont {C.}~\bibnamefont {Lämmerzahl}}, \bibinfo {author}
  {\bibfnamefont {K.}~\bibnamefont {Sengstock}}, \bibinfo {author}
  {\bibfnamefont {N.}~\bibnamefont {Gaaloul}}, \bibinfo {author} {\bibfnamefont
  {W.}~\bibnamefont {Ertmer}}, \ and\ \bibinfo {author} {\bibfnamefont {E.~M.}\
  \bibnamefont {Rasel}},\ }\href {\doibase 10.1088/1367-2630/17/6/065001}
  {\bibfield  {journal} {\bibinfo  {journal} {New Journal of Physics}\ }\textbf
  {\bibinfo {volume} {17}},\ \bibinfo {pages} {065001} (\bibinfo {year}
  {2015})}\BibitemShut {NoStop}%
\bibitem [{\citenamefont {Farkas}\ \emph {et~al.}(2010)\citenamefont {Farkas},
  \citenamefont {Hudek}, \citenamefont {Salim}, \citenamefont {Segal},
  \citenamefont {Squires},\ and\ \citenamefont {Anderson}}]{Farkas10APB}%
  \BibitemOpen
  \bibfield  {author} {\bibinfo {author} {\bibfnamefont {D.~M.}\ \bibnamefont
  {Farkas}}, \bibinfo {author} {\bibfnamefont {K.~M.}\ \bibnamefont {Hudek}},
  \bibinfo {author} {\bibfnamefont {E.~A.}\ \bibnamefont {Salim}}, \bibinfo
  {author} {\bibfnamefont {S.~R.}\ \bibnamefont {Segal}}, \bibinfo {author}
  {\bibfnamefont {M.~B.}\ \bibnamefont {Squires}}, \ and\ \bibinfo {author}
  {\bibfnamefont {D.~Z.}\ \bibnamefont {Anderson}},\ }\href {\doibase
  10.1063/1.3327812} {\bibfield  {journal} {\bibinfo  {journal} {Applied
  Physics Letters}\ }\textbf {\bibinfo {volume} {96}},\ \bibinfo {pages}
  {093102} (\bibinfo {year} {2010})}\BibitemShut {NoStop}%
\bibitem [{\citenamefont {Müntinga}\ \emph {et~al.}(2013)\citenamefont
  {Müntinga}, \citenamefont {Ahlers}, \citenamefont {Krutzik}, \citenamefont
  {Wenzlawski}, \citenamefont {Arnold}, \citenamefont {Becker}, \citenamefont
  {Bongs}, \citenamefont {Dittus}, \citenamefont {Duncker}, \citenamefont
  {Gaaloul}, \citenamefont {Gherasim}, \citenamefont {Giese}, \citenamefont
  {Grzeschik}, \citenamefont {Hänsch}, \citenamefont {Hellmig}, \citenamefont
  {Herr}, \citenamefont {Herrmann}, \citenamefont {Kajari}, \citenamefont
  {Kleinert}, \citenamefont {Lämmerzahl}, \citenamefont {Lewoczko-Adamczyk},
  \citenamefont {Malcolm}, \citenamefont {Meyer}, \citenamefont {Nolte},
  \citenamefont {Peters}, \citenamefont {Popp}, \citenamefont {Reichel},
  \citenamefont {Roura}, \citenamefont {Rudolph}, \citenamefont {Schiemangk},
  \citenamefont {Schneider}, \citenamefont {Seidel}, \citenamefont {Sengstock},
  \citenamefont {Tamma}, \citenamefont {Valenzuela}, \citenamefont {Vogel},
  \citenamefont {Walser}, \citenamefont {Wendrich}, \citenamefont
  {Windpassinger}, \citenamefont {Zeller}, \citenamefont {van Zoest},
  \citenamefont {Ertmer}, \citenamefont {Schleich},\ and\ \citenamefont
  {Rasel}}]{Muentinga13PRL}%
  \BibitemOpen
  \bibfield  {author} {\bibinfo {author} {\bibfnamefont {H.}~\bibnamefont
  {Müntinga}}, \bibinfo {author} {\bibfnamefont {H.}~\bibnamefont {Ahlers}},
  \bibinfo {author} {\bibfnamefont {M.}~\bibnamefont {Krutzik}}, \bibinfo
  {author} {\bibfnamefont {A.}~\bibnamefont {Wenzlawski}}, \bibinfo {author}
  {\bibfnamefont {S.}~\bibnamefont {Arnold}}, \bibinfo {author} {\bibfnamefont
  {D.}~\bibnamefont {Becker}}, \bibinfo {author} {\bibfnamefont
  {K.}~\bibnamefont {Bongs}}, \bibinfo {author} {\bibfnamefont
  {H.}~\bibnamefont {Dittus}}, \bibinfo {author} {\bibfnamefont
  {H.}~\bibnamefont {Duncker}}, \bibinfo {author} {\bibfnamefont
  {N.}~\bibnamefont {Gaaloul}}, \bibinfo {author} {\bibfnamefont
  {C.}~\bibnamefont {Gherasim}}, \bibinfo {author} {\bibfnamefont
  {E.}~\bibnamefont {Giese}}, \bibinfo {author} {\bibfnamefont
  {C.}~\bibnamefont {Grzeschik}}, \bibinfo {author} {\bibfnamefont {T.~W.}\
  \bibnamefont {Hänsch}}, \bibinfo {author} {\bibfnamefont {O.}~\bibnamefont
  {Hellmig}}, \bibinfo {author} {\bibfnamefont {W.}~\bibnamefont {Herr}},
  \bibinfo {author} {\bibfnamefont {S.}~\bibnamefont {Herrmann}}, \bibinfo
  {author} {\bibfnamefont {E.}~\bibnamefont {Kajari}}, \bibinfo {author}
  {\bibfnamefont {S.}~\bibnamefont {Kleinert}}, \bibinfo {author}
  {\bibfnamefont {C.}~\bibnamefont {Lämmerzahl}}, \bibinfo {author}
  {\bibfnamefont {W.}~\bibnamefont {Lewoczko-Adamczyk}}, \bibinfo {author}
  {\bibfnamefont {J.}~\bibnamefont {Malcolm}}, \bibinfo {author} {\bibfnamefont
  {N.}~\bibnamefont {Meyer}}, \bibinfo {author} {\bibfnamefont
  {R.}~\bibnamefont {Nolte}}, \bibinfo {author} {\bibfnamefont
  {A.}~\bibnamefont {Peters}}, \bibinfo {author} {\bibfnamefont
  {M.}~\bibnamefont {Popp}}, \bibinfo {author} {\bibfnamefont {J.}~\bibnamefont
  {Reichel}}, \bibinfo {author} {\bibfnamefont {A.}~\bibnamefont {Roura}},
  \bibinfo {author} {\bibfnamefont {J.}~\bibnamefont {Rudolph}}, \bibinfo
  {author} {\bibfnamefont {M.}~\bibnamefont {Schiemangk}}, \bibinfo {author}
  {\bibfnamefont {M.}~\bibnamefont {Schneider}}, \bibinfo {author}
  {\bibfnamefont {S.~T.}\ \bibnamefont {Seidel}}, \bibinfo {author}
  {\bibfnamefont {K.}~\bibnamefont {Sengstock}}, \bibinfo {author}
  {\bibfnamefont {V.}~\bibnamefont {Tamma}}, \bibinfo {author} {\bibfnamefont
  {T.}~\bibnamefont {Valenzuela}}, \bibinfo {author} {\bibfnamefont
  {A.}~\bibnamefont {Vogel}}, \bibinfo {author} {\bibfnamefont
  {R.}~\bibnamefont {Walser}}, \bibinfo {author} {\bibfnamefont
  {T.}~\bibnamefont {Wendrich}}, \bibinfo {author} {\bibfnamefont
  {P.}~\bibnamefont {Windpassinger}}, \bibinfo {author} {\bibfnamefont
  {W.}~\bibnamefont {Zeller}}, \bibinfo {author} {\bibfnamefont
  {T.}~\bibnamefont {van Zoest}}, \bibinfo {author} {\bibfnamefont
  {W.}~\bibnamefont {Ertmer}}, \bibinfo {author} {\bibfnamefont {W.~P.}\
  \bibnamefont {Schleich}}, \ and\ \bibinfo {author} {\bibfnamefont {E.~M.}\
  \bibnamefont {Rasel}},\ }\href {\doibase 10.1103/physrevlett.110.093602}
  {\bibfield  {journal} {\bibinfo  {journal} {Phys. Rev. Lett.}\ }\textbf
  {\bibinfo {volume} {110}} (\bibinfo {year} {2013}),\
  10.1103/physrevlett.110.093602}\BibitemShut {NoStop}%
\bibitem [{\citenamefont {Lachmann}\ \emph {et~al.}(2021)\citenamefont
  {Lachmann}, \citenamefont {Ahlers}, \citenamefont {Becker}, \citenamefont
  {Dinkelaker}, \citenamefont {Grosse}, \citenamefont {Hellmig}, \citenamefont
  {Müntinga}, \citenamefont {Schkolnik}, \citenamefont {Seidel}, \citenamefont
  {Wendrich}, \citenamefont {Wenzlawski}, \citenamefont {Carrick},
  \citenamefont {Gaaloul}, \citenamefont {Lüdtke}, \citenamefont {Braxmaier},
  \citenamefont {Ertmer}, \citenamefont {Krutzik}, \citenamefont {Lämmerzahl},
  \citenamefont {Peters}, \citenamefont {Schleich}, \citenamefont {Sengstock},
  \citenamefont {Wicht}, \citenamefont {Windpassinger},\ and\ \citenamefont
  {Rasel}}]{Lachmann21NatComm}%
  \BibitemOpen
  \bibfield  {author} {\bibinfo {author} {\bibfnamefont {M.~D.}\ \bibnamefont
  {Lachmann}}, \bibinfo {author} {\bibfnamefont {H.}~\bibnamefont {Ahlers}},
  \bibinfo {author} {\bibfnamefont {D.}~\bibnamefont {Becker}}, \bibinfo
  {author} {\bibfnamefont {A.~N.}\ \bibnamefont {Dinkelaker}}, \bibinfo
  {author} {\bibfnamefont {J.}~\bibnamefont {Grosse}}, \bibinfo {author}
  {\bibfnamefont {O.}~\bibnamefont {Hellmig}}, \bibinfo {author} {\bibfnamefont
  {H.}~\bibnamefont {Müntinga}}, \bibinfo {author} {\bibfnamefont
  {V.}~\bibnamefont {Schkolnik}}, \bibinfo {author} {\bibfnamefont {S.~T.}\
  \bibnamefont {Seidel}}, \bibinfo {author} {\bibfnamefont {T.}~\bibnamefont
  {Wendrich}}, \bibinfo {author} {\bibfnamefont {A.}~\bibnamefont
  {Wenzlawski}}, \bibinfo {author} {\bibfnamefont {B.}~\bibnamefont {Carrick}},
  \bibinfo {author} {\bibfnamefont {N.}~\bibnamefont {Gaaloul}}, \bibinfo
  {author} {\bibfnamefont {D.}~\bibnamefont {Lüdtke}}, \bibinfo {author}
  {\bibfnamefont {C.}~\bibnamefont {Braxmaier}}, \bibinfo {author}
  {\bibfnamefont {W.}~\bibnamefont {Ertmer}}, \bibinfo {author} {\bibfnamefont
  {M.}~\bibnamefont {Krutzik}}, \bibinfo {author} {\bibfnamefont
  {C.}~\bibnamefont {Lämmerzahl}}, \bibinfo {author} {\bibfnamefont
  {A.}~\bibnamefont {Peters}}, \bibinfo {author} {\bibfnamefont {W.~P.}\
  \bibnamefont {Schleich}}, \bibinfo {author} {\bibfnamefont {K.}~\bibnamefont
  {Sengstock}}, \bibinfo {author} {\bibfnamefont {A.}~\bibnamefont {Wicht}},
  \bibinfo {author} {\bibfnamefont {P.}~\bibnamefont {Windpassinger}}, \ and\
  \bibinfo {author} {\bibfnamefont {E.~M.}\ \bibnamefont {Rasel}},\ }\href
  {\doibase 10.1038/s41467-021-21628-z} {\bibfield  {journal} {\bibinfo
  {journal} {Nature Communications}\ }\textbf {\bibinfo {volume} {12}},\
  \bibinfo {pages} {1317} (\bibinfo {year} {2021})}\BibitemShut {NoStop}%
\bibitem [{\citenamefont {Deppner}\ \emph {et~al.}(2021)\citenamefont
  {Deppner}, \citenamefont {Herr}, \citenamefont {Cornelius}, \citenamefont
  {Stromberger}, \citenamefont {Sternke}, \citenamefont {Grzeschik},
  \citenamefont {Grote}, \citenamefont {Rudolph}, \citenamefont {Herrmann},
  \citenamefont {Krutzik}, \citenamefont {Wenzlawski}, \citenamefont {Corgier},
  \citenamefont {Charron}, \citenamefont {Gu\'ery-Odelin}, \citenamefont
  {Gaaloul}, \citenamefont {L\"ammerzahl}, \citenamefont {Peters},
  \citenamefont {Windpassinger},\ and\ \citenamefont {Rasel}}]{Deppner21PRL}%
  \BibitemOpen
  \bibfield  {author} {\bibinfo {author} {\bibfnamefont {C.}~\bibnamefont
  {Deppner}}, \bibinfo {author} {\bibfnamefont {W.}~\bibnamefont {Herr}},
  \bibinfo {author} {\bibfnamefont {M.}~\bibnamefont {Cornelius}}, \bibinfo
  {author} {\bibfnamefont {P.}~\bibnamefont {Stromberger}}, \bibinfo {author}
  {\bibfnamefont {T.}~\bibnamefont {Sternke}}, \bibinfo {author} {\bibfnamefont
  {C.}~\bibnamefont {Grzeschik}}, \bibinfo {author} {\bibfnamefont
  {A.}~\bibnamefont {Grote}}, \bibinfo {author} {\bibfnamefont
  {J.}~\bibnamefont {Rudolph}}, \bibinfo {author} {\bibfnamefont
  {S.}~\bibnamefont {Herrmann}}, \bibinfo {author} {\bibfnamefont
  {M.}~\bibnamefont {Krutzik}}, \bibinfo {author} {\bibfnamefont
  {A.}~\bibnamefont {Wenzlawski}}, \bibinfo {author} {\bibfnamefont
  {R.}~\bibnamefont {Corgier}}, \bibinfo {author} {\bibfnamefont
  {E.}~\bibnamefont {Charron}}, \bibinfo {author} {\bibfnamefont
  {D.}~\bibnamefont {Gu\'ery-Odelin}}, \bibinfo {author} {\bibfnamefont
  {N.}~\bibnamefont {Gaaloul}}, \bibinfo {author} {\bibfnamefont
  {C.}~\bibnamefont {L\"ammerzahl}}, \bibinfo {author} {\bibfnamefont
  {A.}~\bibnamefont {Peters}}, \bibinfo {author} {\bibfnamefont
  {P.}~\bibnamefont {Windpassinger}}, \ and\ \bibinfo {author} {\bibfnamefont
  {E.~M.}\ \bibnamefont {Rasel}},\ }\href {\doibase
  10.1103/PhysRevLett.127.100401} {\bibfield  {journal} {\bibinfo  {journal}
  {Phys. Rev. Lett.}\ }\textbf {\bibinfo {volume} {127}},\ \bibinfo {pages}
  {100401} (\bibinfo {year} {2021})}\BibitemShut {NoStop}%
\bibitem [{\citenamefont {Becker}\ \emph {et~al.}(2018)\citenamefont {Becker},
  \citenamefont {Lachmann}, \citenamefont {Seidel}, \citenamefont {Ahlers},
  \citenamefont {Dinkelaker}, \citenamefont {Grosse}, \citenamefont {Hellmig},
  \citenamefont {Müntinga}, \citenamefont {Schkolnik}, \citenamefont
  {Wendrich}, \citenamefont {Wenzlawski}, \citenamefont {Weps}, \citenamefont
  {Corgier}, \citenamefont {Franz}, \citenamefont {Gaaloul}, \citenamefont
  {Herr}, \citenamefont {Lüdtke}, \citenamefont {Popp}, \citenamefont {Amri},
  \citenamefont {Duncker}, \citenamefont {Erbe}, \citenamefont {Kohfeldt},
  \citenamefont {Kubelka-Lange}, \citenamefont {Braxmaier}, \citenamefont
  {Charron}, \citenamefont {Ertmer}, \citenamefont {Krutzik}, \citenamefont
  {Lämmerzahl}, \citenamefont {Peters}, \citenamefont {Schleich},
  \citenamefont {Sengstock}, \citenamefont {Walser}, \citenamefont {Wicht},
  \citenamefont {Windpassinger},\ and\ \citenamefont {Rasel}}]{Becker2018}%
  \BibitemOpen
  \bibfield  {author} {\bibinfo {author} {\bibfnamefont {D.}~\bibnamefont
  {Becker}}, \bibinfo {author} {\bibfnamefont {M.~D.}\ \bibnamefont
  {Lachmann}}, \bibinfo {author} {\bibfnamefont {S.~T.}\ \bibnamefont
  {Seidel}}, \bibinfo {author} {\bibfnamefont {H.}~\bibnamefont {Ahlers}},
  \bibinfo {author} {\bibfnamefont {A.~N.}\ \bibnamefont {Dinkelaker}},
  \bibinfo {author} {\bibfnamefont {J.}~\bibnamefont {Grosse}}, \bibinfo
  {author} {\bibfnamefont {O.}~\bibnamefont {Hellmig}}, \bibinfo {author}
  {\bibfnamefont {H.}~\bibnamefont {Müntinga}}, \bibinfo {author}
  {\bibfnamefont {V.}~\bibnamefont {Schkolnik}}, \bibinfo {author}
  {\bibfnamefont {T.}~\bibnamefont {Wendrich}}, \bibinfo {author}
  {\bibfnamefont {A.}~\bibnamefont {Wenzlawski}}, \bibinfo {author}
  {\bibfnamefont {B.}~\bibnamefont {Weps}}, \bibinfo {author} {\bibfnamefont
  {R.}~\bibnamefont {Corgier}}, \bibinfo {author} {\bibfnamefont
  {T.}~\bibnamefont {Franz}}, \bibinfo {author} {\bibfnamefont
  {N.}~\bibnamefont {Gaaloul}}, \bibinfo {author} {\bibfnamefont
  {W.}~\bibnamefont {Herr}}, \bibinfo {author} {\bibfnamefont {D.}~\bibnamefont
  {Lüdtke}}, \bibinfo {author} {\bibfnamefont {M.}~\bibnamefont {Popp}},
  \bibinfo {author} {\bibfnamefont {S.}~\bibnamefont {Amri}}, \bibinfo {author}
  {\bibfnamefont {H.}~\bibnamefont {Duncker}}, \bibinfo {author} {\bibfnamefont
  {M.}~\bibnamefont {Erbe}}, \bibinfo {author} {\bibfnamefont {A.}~\bibnamefont
  {Kohfeldt}}, \bibinfo {author} {\bibfnamefont {A.}~\bibnamefont
  {Kubelka-Lange}}, \bibinfo {author} {\bibfnamefont {C.}~\bibnamefont
  {Braxmaier}}, \bibinfo {author} {\bibfnamefont {E.}~\bibnamefont {Charron}},
  \bibinfo {author} {\bibfnamefont {W.}~\bibnamefont {Ertmer}}, \bibinfo
  {author} {\bibfnamefont {M.}~\bibnamefont {Krutzik}}, \bibinfo {author}
  {\bibfnamefont {C.}~\bibnamefont {Lämmerzahl}}, \bibinfo {author}
  {\bibfnamefont {A.}~\bibnamefont {Peters}}, \bibinfo {author} {\bibfnamefont
  {W.~P.}\ \bibnamefont {Schleich}}, \bibinfo {author} {\bibfnamefont
  {K.}~\bibnamefont {Sengstock}}, \bibinfo {author} {\bibfnamefont
  {R.}~\bibnamefont {Walser}}, \bibinfo {author} {\bibfnamefont
  {A.}~\bibnamefont {Wicht}}, \bibinfo {author} {\bibfnamefont
  {P.}~\bibnamefont {Windpassinger}}, \ and\ \bibinfo {author} {\bibfnamefont
  {E.~M.}\ \bibnamefont {Rasel}},\ }\href {\doibase 10.1038/s41586-018-0605-1}
  {\bibfield  {journal} {\bibinfo  {journal} {Nature}\ }\textbf {\bibinfo
  {volume} {562}},\ \bibinfo {pages} {391} (\bibinfo {year}
  {2018})}\BibitemShut {NoStop}%
\bibitem [{\citenamefont {Aveline}\ \emph {et~al.}(2020)\citenamefont
  {Aveline}, \citenamefont {Williams}, \citenamefont {Elliott}, \citenamefont
  {Dutenhoffer}, \citenamefont {Kellogg}, \citenamefont {Kohel}, \citenamefont
  {Lay}, \citenamefont {Oudrhiri}, \citenamefont {Shotwell}, \citenamefont
  {Yu},\ and\ \citenamefont {Thompson}}]{Aveline20Nature}%
  \BibitemOpen
  \bibfield  {author} {\bibinfo {author} {\bibfnamefont {D.~C.}\ \bibnamefont
  {Aveline}}, \bibinfo {author} {\bibfnamefont {J.~R.}\ \bibnamefont
  {Williams}}, \bibinfo {author} {\bibfnamefont {E.~R.}\ \bibnamefont
  {Elliott}}, \bibinfo {author} {\bibfnamefont {C.}~\bibnamefont
  {Dutenhoffer}}, \bibinfo {author} {\bibfnamefont {J.~R.}\ \bibnamefont
  {Kellogg}}, \bibinfo {author} {\bibfnamefont {J.~M.}\ \bibnamefont {Kohel}},
  \bibinfo {author} {\bibfnamefont {N.~E.}\ \bibnamefont {Lay}}, \bibinfo
  {author} {\bibfnamefont {K.}~\bibnamefont {Oudrhiri}}, \bibinfo {author}
  {\bibfnamefont {R.~F.}\ \bibnamefont {Shotwell}}, \bibinfo {author}
  {\bibfnamefont {N.}~\bibnamefont {Yu}}, \ and\ \bibinfo {author}
  {\bibfnamefont {R.~J.}\ \bibnamefont {Thompson}},\ }\href {\doibase
  10.1038/s41586-020-2346-1} {\bibfield  {journal} {\bibinfo  {journal}
  {Nature}\ }\textbf {\bibinfo {volume} {582}},\ \bibinfo {pages} {193}
  (\bibinfo {year} {2020})}\BibitemShut {NoStop}%
\bibitem [{\citenamefont {Frye}\ \emph {et~al.}(2021)\citenamefont {Frye},
  \citenamefont {Abend}, \citenamefont {Bartosch}, \citenamefont {Bawamia},
  \citenamefont {Becker}, \citenamefont {Blume}, \citenamefont {Braxmaier},
  \citenamefont {Chiow}, \citenamefont {Efremov}, \citenamefont {Ertmer},
  \citenamefont {Fierlinger}, \citenamefont {Franz}, \citenamefont {Gaaloul},
  \citenamefont {Grosse}, \citenamefont {Grzeschik}, \citenamefont {Hellmig},
  \citenamefont {Henderson}, \citenamefont {Herr}, \citenamefont {Israelsson},
  \citenamefont {Kohel}, \citenamefont {Krutzik}, \citenamefont {Kürbis},
  \citenamefont {L{\"a}mmerzahl}, \citenamefont {List}, \citenamefont
  {Lüdtke}, \citenamefont {Lundblad}, \citenamefont {Marburger}, \citenamefont
  {Meister}, \citenamefont {Mihm}, \citenamefont {Müller}, \citenamefont
  {Müntinga}, \citenamefont {Nepal}, \citenamefont {Oberschulte},
  \citenamefont {Papakonstantinou}, \citenamefont {Perovs\u{s}ek},
  \citenamefont {Peters}, \citenamefont {Prat}, \citenamefont {Rasel},
  \citenamefont {Roura}, \citenamefont {Sbroscia}, \citenamefont {Schleich},
  \citenamefont {Schubert}, \citenamefont {Seidel}, \citenamefont {Sommer},
  \citenamefont {Spindeldreier}, \citenamefont {Stamper-Kurn}, \citenamefont
  {Stuhl}, \citenamefont {Warner}, \citenamefont {Wendrich}, \citenamefont
  {Wenzlawski}, \citenamefont {Wicht}, \citenamefont {Windpassinger},
  \citenamefont {Yu},\ and\ \citenamefont {Wörner}}]{Frye2021EPJQT}%
  \BibitemOpen
  \bibfield  {author} {\bibinfo {author} {\bibfnamefont {K.}~\bibnamefont
  {Frye}}, \bibinfo {author} {\bibfnamefont {S.}~\bibnamefont {Abend}},
  \bibinfo {author} {\bibfnamefont {W.}~\bibnamefont {Bartosch}}, \bibinfo
  {author} {\bibfnamefont {A.}~\bibnamefont {Bawamia}}, \bibinfo {author}
  {\bibfnamefont {D.}~\bibnamefont {Becker}}, \bibinfo {author} {\bibfnamefont
  {H.}~\bibnamefont {Blume}}, \bibinfo {author} {\bibfnamefont
  {C.}~\bibnamefont {Braxmaier}}, \bibinfo {author} {\bibfnamefont {S.-W.}\
  \bibnamefont {Chiow}}, \bibinfo {author} {\bibfnamefont {M.~A.}\ \bibnamefont
  {Efremov}}, \bibinfo {author} {\bibfnamefont {W.}~\bibnamefont {Ertmer}},
  \bibinfo {author} {\bibfnamefont {P.}~\bibnamefont {Fierlinger}}, \bibinfo
  {author} {\bibfnamefont {T.}~\bibnamefont {Franz}}, \bibinfo {author}
  {\bibfnamefont {N.}~\bibnamefont {Gaaloul}}, \bibinfo {author} {\bibfnamefont
  {J.}~\bibnamefont {Grosse}}, \bibinfo {author} {\bibfnamefont
  {C.}~\bibnamefont {Grzeschik}}, \bibinfo {author} {\bibfnamefont
  {O.}~\bibnamefont {Hellmig}}, \bibinfo {author} {\bibfnamefont {V.~A.}\
  \bibnamefont {Henderson}}, \bibinfo {author} {\bibfnamefont {W.}~\bibnamefont
  {Herr}}, \bibinfo {author} {\bibfnamefont {U.}~\bibnamefont {Israelsson}},
  \bibinfo {author} {\bibfnamefont {J.}~\bibnamefont {Kohel}}, \bibinfo
  {author} {\bibfnamefont {M.}~\bibnamefont {Krutzik}}, \bibinfo {author}
  {\bibfnamefont {C.}~\bibnamefont {Kürbis}}, \bibinfo {author} {\bibfnamefont
  {C.}~\bibnamefont {L{\"a}mmerzahl}}, \bibinfo {author} {\bibfnamefont
  {M.}~\bibnamefont {List}}, \bibinfo {author} {\bibfnamefont {D.}~\bibnamefont
  {Lüdtke}}, \bibinfo {author} {\bibfnamefont {N.}~\bibnamefont {Lundblad}},
  \bibinfo {author} {\bibfnamefont {J.~P.}\ \bibnamefont {Marburger}}, \bibinfo
  {author} {\bibfnamefont {M.}~\bibnamefont {Meister}}, \bibinfo {author}
  {\bibfnamefont {M.}~\bibnamefont {Mihm}}, \bibinfo {author} {\bibfnamefont
  {H.}~\bibnamefont {Müller}}, \bibinfo {author} {\bibfnamefont
  {H.}~\bibnamefont {Müntinga}}, \bibinfo {author} {\bibfnamefont {A.~M.}\
  \bibnamefont {Nepal}}, \bibinfo {author} {\bibfnamefont {T.}~\bibnamefont
  {Oberschulte}}, \bibinfo {author} {\bibfnamefont {A.}~\bibnamefont
  {Papakonstantinou}}, \bibinfo {author} {\bibfnamefont {J.}~\bibnamefont
  {Perovs\u{s}ek}}, \bibinfo {author} {\bibfnamefont {A.}~\bibnamefont
  {Peters}}, \bibinfo {author} {\bibfnamefont {A.}~\bibnamefont {Prat}},
  \bibinfo {author} {\bibfnamefont {E.~M.}\ \bibnamefont {Rasel}}, \bibinfo
  {author} {\bibfnamefont {A.}~\bibnamefont {Roura}}, \bibinfo {author}
  {\bibfnamefont {M.}~\bibnamefont {Sbroscia}}, \bibinfo {author}
  {\bibfnamefont {W.~P.}\ \bibnamefont {Schleich}}, \bibinfo {author}
  {\bibfnamefont {C.}~\bibnamefont {Schubert}}, \bibinfo {author}
  {\bibfnamefont {S.~T.}\ \bibnamefont {Seidel}}, \bibinfo {author}
  {\bibfnamefont {J.}~\bibnamefont {Sommer}}, \bibinfo {author} {\bibfnamefont
  {C.}~\bibnamefont {Spindeldreier}}, \bibinfo {author} {\bibfnamefont
  {D.}~\bibnamefont {Stamper-Kurn}}, \bibinfo {author} {\bibfnamefont {B.~K.}\
  \bibnamefont {Stuhl}}, \bibinfo {author} {\bibfnamefont {M.}~\bibnamefont
  {Warner}}, \bibinfo {author} {\bibfnamefont {T.}~\bibnamefont {Wendrich}},
  \bibinfo {author} {\bibfnamefont {A.}~\bibnamefont {Wenzlawski}}, \bibinfo
  {author} {\bibfnamefont {A.}~\bibnamefont {Wicht}}, \bibinfo {author}
  {\bibfnamefont {P.}~\bibnamefont {Windpassinger}}, \bibinfo {author}
  {\bibfnamefont {N.}~\bibnamefont {Yu}}, \ and\ \bibinfo {author}
  {\bibfnamefont {L.}~\bibnamefont {Wörner}},\ }\href {\doibase
  10.1140/epjqt/s40507-020-00090-8} {\bibfield  {journal} {\bibinfo  {journal}
  {{EPJ} Quantum Technology}\ }\textbf {\bibinfo {volume} {8}} (\bibinfo {year}
  {2021}),\ 10.1140/epjqt/s40507-020-00090-8}\BibitemShut {NoStop}%
\bibitem [{\citenamefont {Haslinger}\ \emph {et~al.}(2017)\citenamefont
  {Haslinger}, \citenamefont {Jaffe}, \citenamefont {Xu}, \citenamefont
  {Schwartz}, \citenamefont {Sonnleitner}, \citenamefont {Ritsch-Marte},
  \citenamefont {Ritsch},\ and\ \citenamefont
  {Müller}}]{HaslingerNaturePhys2017}%
  \BibitemOpen
  \bibfield  {author} {\bibinfo {author} {\bibfnamefont {P.}~\bibnamefont
  {Haslinger}}, \bibinfo {author} {\bibfnamefont {M.}~\bibnamefont {Jaffe}},
  \bibinfo {author} {\bibfnamefont {V.}~\bibnamefont {Xu}}, \bibinfo {author}
  {\bibfnamefont {O.}~\bibnamefont {Schwartz}}, \bibinfo {author}
  {\bibfnamefont {M.}~\bibnamefont {Sonnleitner}}, \bibinfo {author}
  {\bibfnamefont {M.}~\bibnamefont {Ritsch-Marte}}, \bibinfo {author}
  {\bibfnamefont {H.}~\bibnamefont {Ritsch}}, \ and\ \bibinfo {author}
  {\bibfnamefont {H.}~\bibnamefont {Müller}},\ }\href {\doibase
  10.1038/s41567-017-0004-9} {\bibfield  {journal} {\bibinfo  {journal} {Nature
  Physics}\ }\textbf {\bibinfo {volume} {14}},\ \bibinfo {pages} {257}
  (\bibinfo {year} {2017})}\BibitemShut {NoStop}%
\bibitem [{\citenamefont {Barrett}\ \emph {et~al.}(2001)\citenamefont
  {Barrett}, \citenamefont {Sauer},\ and\ \citenamefont
  {Chapman}}]{Barrett01PRL}%
  \BibitemOpen
  \bibfield  {author} {\bibinfo {author} {\bibfnamefont {M.~D.}\ \bibnamefont
  {Barrett}}, \bibinfo {author} {\bibfnamefont {J.~A.}\ \bibnamefont {Sauer}},
  \ and\ \bibinfo {author} {\bibfnamefont {M.~S.}\ \bibnamefont {Chapman}},\
  }\href {\doibase 10.1103/physrevlett.87.010404} {\bibfield  {journal}
  {\bibinfo  {journal} {Physical Review Letters}\ }\textbf {\bibinfo {volume}
  {87}} (\bibinfo {year} {2001}),\ 10.1103/physrevlett.87.010404}\BibitemShut
  {NoStop}%
\bibitem [{\citenamefont {Cl\'ement}\ \emph {et~al.}(2009)\citenamefont
  {Cl\'ement}, \citenamefont {Brantut}, \citenamefont {Robert-de
  Saint-Vincent}, \citenamefont {Nyman}, \citenamefont {Aspect}, \citenamefont
  {Bourdel},\ and\ \citenamefont {Bouyer}}]{Clement09PRAR}%
  \BibitemOpen
  \bibfield  {author} {\bibinfo {author} {\bibfnamefont {J.-F.}\ \bibnamefont
  {Cl\'ement}}, \bibinfo {author} {\bibfnamefont {J.-P.}\ \bibnamefont
  {Brantut}}, \bibinfo {author} {\bibfnamefont {M.}~\bibnamefont {Robert-de
  Saint-Vincent}}, \bibinfo {author} {\bibfnamefont {R.~A.}\ \bibnamefont
  {Nyman}}, \bibinfo {author} {\bibfnamefont {A.}~\bibnamefont {Aspect}},
  \bibinfo {author} {\bibfnamefont {T.}~\bibnamefont {Bourdel}}, \ and\
  \bibinfo {author} {\bibfnamefont {P.}~\bibnamefont {Bouyer}},\ }\href
  {\doibase 10.1103/PhysRevA.79.061406} {\bibfield  {journal} {\bibinfo
  {journal} {Phys. Rev. A}\ }\textbf {\bibinfo {volume} {79}},\ \bibinfo
  {pages} {061406} (\bibinfo {year} {2009})}\BibitemShut {NoStop}%
\bibitem [{\citenamefont {Landini}\ \emph {et~al.}(2012)\citenamefont
  {Landini}, \citenamefont {Roy}, \citenamefont {Roati}, \citenamefont
  {Simoni}, \citenamefont {Inguscio}, \citenamefont {Modugno},\ and\
  \citenamefont {Fattori}}]{Landini12PRA}%
  \BibitemOpen
  \bibfield  {author} {\bibinfo {author} {\bibfnamefont {M.}~\bibnamefont
  {Landini}}, \bibinfo {author} {\bibfnamefont {S.}~\bibnamefont {Roy}},
  \bibinfo {author} {\bibfnamefont {G.}~\bibnamefont {Roati}}, \bibinfo
  {author} {\bibfnamefont {A.}~\bibnamefont {Simoni}}, \bibinfo {author}
  {\bibfnamefont {M.}~\bibnamefont {Inguscio}}, \bibinfo {author}
  {\bibfnamefont {G.}~\bibnamefont {Modugno}}, \ and\ \bibinfo {author}
  {\bibfnamefont {M.}~\bibnamefont {Fattori}},\ }\href {\doibase
  10.1103/physreva.86.033421} {\bibfield  {journal} {\bibinfo  {journal}
  {Physical Review A}\ }\textbf {\bibinfo {volume} {86}} (\bibinfo {year}
  {2012}),\ 10.1103/physreva.86.033421}\BibitemShut {NoStop}%
\bibitem [{\citenamefont {Stellmer}\ \emph {et~al.}(2013)\citenamefont
  {Stellmer}, \citenamefont {Grimm},\ and\ \citenamefont
  {Schreck}}]{Stellmer13PRA}%
  \BibitemOpen
  \bibfield  {author} {\bibinfo {author} {\bibfnamefont {S.}~\bibnamefont
  {Stellmer}}, \bibinfo {author} {\bibfnamefont {R.}~\bibnamefont {Grimm}}, \
  and\ \bibinfo {author} {\bibfnamefont {F.}~\bibnamefont {Schreck}},\ }\href
  {\doibase 10.1103/physreva.87.013611} {\bibfield  {journal} {\bibinfo
  {journal} {Physical Review A}\ }\textbf {\bibinfo {volume} {87}} (\bibinfo
  {year} {2013}),\ 10.1103/physreva.87.013611}\BibitemShut {NoStop}%
\bibitem [{\citenamefont {Kulas}\ \emph {et~al.}(2016)\citenamefont {Kulas},
  \citenamefont {Vogt}, \citenamefont {Resch}, \citenamefont {Hartwig},
  \citenamefont {Ganske}, \citenamefont {Matthias}, \citenamefont {Schlippert},
  \citenamefont {Wendrich}, \citenamefont {Ertmer}, \citenamefont {Rasel},
  \citenamefont {Damjanic}, \citenamefont {We{\ss}els}, \citenamefont
  {Kohfeldt}, \citenamefont {Luvsandamdin}, \citenamefont {Schiemangk},
  \citenamefont {Grzeschik}, \citenamefont {Krutzik}, \citenamefont {Wicht},
  \citenamefont {Peters}, \citenamefont {Herrmann},\ and\ \citenamefont
  {Lämmerzahl}}]{Kulas2016}%
  \BibitemOpen
  \bibfield  {author} {\bibinfo {author} {\bibfnamefont {S.}~\bibnamefont
  {Kulas}}, \bibinfo {author} {\bibfnamefont {C.}~\bibnamefont {Vogt}},
  \bibinfo {author} {\bibfnamefont {A.}~\bibnamefont {Resch}}, \bibinfo
  {author} {\bibfnamefont {J.}~\bibnamefont {Hartwig}}, \bibinfo {author}
  {\bibfnamefont {S.}~\bibnamefont {Ganske}}, \bibinfo {author} {\bibfnamefont
  {J.}~\bibnamefont {Matthias}}, \bibinfo {author} {\bibfnamefont
  {D.}~\bibnamefont {Schlippert}}, \bibinfo {author} {\bibfnamefont
  {T.}~\bibnamefont {Wendrich}}, \bibinfo {author} {\bibfnamefont
  {W.}~\bibnamefont {Ertmer}}, \bibinfo {author} {\bibfnamefont {E.~M.}\
  \bibnamefont {Rasel}}, \bibinfo {author} {\bibfnamefont {M.}~\bibnamefont
  {Damjanic}}, \bibinfo {author} {\bibfnamefont {P.}~\bibnamefont
  {We{\ss}els}}, \bibinfo {author} {\bibfnamefont {A.}~\bibnamefont
  {Kohfeldt}}, \bibinfo {author} {\bibfnamefont {E.}~\bibnamefont
  {Luvsandamdin}}, \bibinfo {author} {\bibfnamefont {M.}~\bibnamefont
  {Schiemangk}}, \bibinfo {author} {\bibfnamefont {C.}~\bibnamefont
  {Grzeschik}}, \bibinfo {author} {\bibfnamefont {M.}~\bibnamefont {Krutzik}},
  \bibinfo {author} {\bibfnamefont {A.}~\bibnamefont {Wicht}}, \bibinfo
  {author} {\bibfnamefont {A.}~\bibnamefont {Peters}}, \bibinfo {author}
  {\bibfnamefont {S.}~\bibnamefont {Herrmann}}, \ and\ \bibinfo {author}
  {\bibfnamefont {C.}~\bibnamefont {Lämmerzahl}},\ }\href {\doibase
  10.1007/s12217-016-9524-7} {\bibfield  {journal} {\bibinfo  {journal}
  {Microgravity Science and Technology}\ }\textbf {\bibinfo {volume} {29}},\
  \bibinfo {pages} {37} (\bibinfo {year} {2016})}\BibitemShut {NoStop}%
\bibitem [{\citenamefont {Vogt}\ \emph {et~al.}(2020)\citenamefont {Vogt},
  \citenamefont {Woltmann}, \citenamefont {Herrmann}, \citenamefont
  {Lämmerzahl}, \citenamefont {Albers}, \citenamefont {Schlippert},\ and\
  \citenamefont {and}}]{Vogt2019}%
  \BibitemOpen
  \bibfield  {author} {\bibinfo {author} {\bibfnamefont {C.}~\bibnamefont
  {Vogt}}, \bibinfo {author} {\bibfnamefont {M.}~\bibnamefont {Woltmann}},
  \bibinfo {author} {\bibfnamefont {S.}~\bibnamefont {Herrmann}}, \bibinfo
  {author} {\bibfnamefont {C.}~\bibnamefont {Lämmerzahl}}, \bibinfo {author}
  {\bibfnamefont {H.}~\bibnamefont {Albers}}, \bibinfo {author} {\bibfnamefont
  {D.}~\bibnamefont {Schlippert}}, \ and\ \bibinfo {author} {\bibfnamefont
  {E.~M.~R.}\ \bibnamefont {and}},\ }\href {\doibase
  10.1103/physreva.101.013634} {\bibfield  {journal} {\bibinfo  {journal}
  {Physical Review A}\ }\textbf {\bibinfo {volume} {101}},\ \bibinfo {pages}
  {013634} (\bibinfo {year} {2020})}\BibitemShut {NoStop}%
\bibitem [{\citenamefont {O'Hara}\ \emph {et~al.}(2001)\citenamefont {O'Hara},
  \citenamefont {Gehm}, \citenamefont {Granade},\ and\ \citenamefont
  {Thomas}}]{OHara01PRA}%
  \BibitemOpen
  \bibfield  {author} {\bibinfo {author} {\bibfnamefont {K.~M.}\ \bibnamefont
  {O'Hara}}, \bibinfo {author} {\bibfnamefont {M.~E.}\ \bibnamefont {Gehm}},
  \bibinfo {author} {\bibfnamefont {S.~R.}\ \bibnamefont {Granade}}, \ and\
  \bibinfo {author} {\bibfnamefont {J.~E.}\ \bibnamefont {Thomas}},\ }\href
  {\doibase 10.1103/physreva.64.051403} {\bibfield  {journal} {\bibinfo
  {journal} {Physical Review A}\ }\textbf {\bibinfo {volume} {64}},\ \bibinfo
  {pages} {051403} (\bibinfo {year} {2001})}\BibitemShut {NoStop}%
\bibitem [{\citenamefont {Kinoshita}\ \emph {et~al.}(2005)\citenamefont
  {Kinoshita}, \citenamefont {Wenger},\ and\ \citenamefont
  {Weiss}}]{Kinoshita05PRAR}%
  \BibitemOpen
  \bibfield  {author} {\bibinfo {author} {\bibfnamefont {T.}~\bibnamefont
  {Kinoshita}}, \bibinfo {author} {\bibfnamefont {T.}~\bibnamefont {Wenger}}, \
  and\ \bibinfo {author} {\bibfnamefont {D.}~\bibnamefont {Weiss}},\ }\href
  {\doibase 10.1103/PhysRevA.71.011602} {\bibfield  {journal} {\bibinfo
  {journal} {Phys. Rev. A}\ }\textbf {\bibinfo {volume} {71}},\ \bibinfo
  {pages} {011602} (\bibinfo {year} {2005})}\BibitemShut {NoStop}%
\bibitem [{\citenamefont {Roy}\ \emph {et~al.}(2016)\citenamefont {Roy},
  \citenamefont {Green}, \citenamefont {Bowler},\ and\ \citenamefont
  {Gupta}}]{Roy2016PRA}%
  \BibitemOpen
  \bibfield  {author} {\bibinfo {author} {\bibfnamefont {R.}~\bibnamefont
  {Roy}}, \bibinfo {author} {\bibfnamefont {A.}~\bibnamefont {Green}}, \bibinfo
  {author} {\bibfnamefont {R.}~\bibnamefont {Bowler}}, \ and\ \bibinfo {author}
  {\bibfnamefont {S.}~\bibnamefont {Gupta}},\ }\href {\doibase
  10.1103/PhysRevA.93.043403} {\bibfield  {journal} {\bibinfo  {journal} {Phys.
  Rev. A}\ }\textbf {\bibinfo {volume} {93}},\ \bibinfo {pages} {043403}
  (\bibinfo {year} {2016})}\BibitemShut {NoStop}%
\bibitem [{\citenamefont {Condon}\ \emph {et~al.}(2019)\citenamefont {Condon},
  \citenamefont {Rabault}, \citenamefont {Barrett}, \citenamefont {Chichet},
  \citenamefont {Arguel}, \citenamefont {Eneriz-Imaz}, \citenamefont {Naik},
  \citenamefont {Bertoldi}, \citenamefont {Battelier}, \citenamefont {Bouyer},\
  and\ \citenamefont {Landragin}}]{Condon2019}%
  \BibitemOpen
  \bibfield  {author} {\bibinfo {author} {\bibfnamefont {G.}~\bibnamefont
  {Condon}}, \bibinfo {author} {\bibfnamefont {M.}~\bibnamefont {Rabault}},
  \bibinfo {author} {\bibfnamefont {B.}~\bibnamefont {Barrett}}, \bibinfo
  {author} {\bibfnamefont {L.}~\bibnamefont {Chichet}}, \bibinfo {author}
  {\bibfnamefont {R.}~\bibnamefont {Arguel}}, \bibinfo {author} {\bibfnamefont
  {H.}~\bibnamefont {Eneriz-Imaz}}, \bibinfo {author} {\bibfnamefont
  {D.}~\bibnamefont {Naik}}, \bibinfo {author} {\bibfnamefont {A.}~\bibnamefont
  {Bertoldi}}, \bibinfo {author} {\bibfnamefont {B.}~\bibnamefont {Battelier}},
  \bibinfo {author} {\bibfnamefont {P.}~\bibnamefont {Bouyer}}, \ and\ \bibinfo
  {author} {\bibfnamefont {A.}~\bibnamefont {Landragin}},\ }\href {\doibase
  10.1103/physrevlett.123.240402} {\bibfield  {journal} {\bibinfo  {journal}
  {Physical Review Letters}\ }\textbf {\bibinfo {volume} {123}} (\bibinfo
  {year} {2019}),\ 10.1103/physrevlett.123.240402}\BibitemShut {NoStop}%
\bibitem [{\citenamefont {Albers}\ \emph {et~al.}(2022)\citenamefont {Albers},
  \citenamefont {Corgier}, \citenamefont {Herbst}, \citenamefont {Rajagopalan},
  \citenamefont {Schubert}, \citenamefont {Vogt}, \citenamefont {Woltmann},
  \citenamefont {Lämmerzahl}, \citenamefont {Herrmann}, \citenamefont
  {Charron}, \citenamefont {Ertmer}, \citenamefont {Rasel}, \citenamefont
  {Gaaloul},\ and\ \citenamefont {Schlippert}}]{Albers2022Commun}%
  \BibitemOpen
  \bibfield  {author} {\bibinfo {author} {\bibfnamefont {H.}~\bibnamefont
  {Albers}}, \bibinfo {author} {\bibfnamefont {R.}~\bibnamefont {Corgier}},
  \bibinfo {author} {\bibfnamefont {A.}~\bibnamefont {Herbst}}, \bibinfo
  {author} {\bibfnamefont {A.}~\bibnamefont {Rajagopalan}}, \bibinfo {author}
  {\bibfnamefont {C.}~\bibnamefont {Schubert}}, \bibinfo {author}
  {\bibfnamefont {C.}~\bibnamefont {Vogt}}, \bibinfo {author} {\bibfnamefont
  {M.}~\bibnamefont {Woltmann}}, \bibinfo {author} {\bibfnamefont
  {C.}~\bibnamefont {Lämmerzahl}}, \bibinfo {author} {\bibfnamefont
  {S.}~\bibnamefont {Herrmann}}, \bibinfo {author} {\bibfnamefont
  {E.}~\bibnamefont {Charron}}, \bibinfo {author} {\bibfnamefont
  {W.}~\bibnamefont {Ertmer}}, \bibinfo {author} {\bibfnamefont {E.~M.}\
  \bibnamefont {Rasel}}, \bibinfo {author} {\bibfnamefont {N.}~\bibnamefont
  {Gaaloul}}, \ and\ \bibinfo {author} {\bibfnamefont {D.}~\bibnamefont
  {Schlippert}},\ }\href {\doibase 10.1038/s42005-022-00825-2} {\bibfield
  {journal} {\bibinfo  {journal} {Communications Physics}\ }\textbf {\bibinfo
  {volume} {5}},\ \bibinfo {pages} {60} (\bibinfo {year} {2022})}\BibitemShut
  {NoStop}%
\bibitem [{\citenamefont {Inouye}\ \emph {et~al.}(1998)\citenamefont {Inouye},
  \citenamefont {Andrews}, \citenamefont {Stenger}, \citenamefont {Miesner},
  \citenamefont {Stamper-Kurn},\ and\ \citenamefont
  {Ketterle}}]{Inouye98Nature}%
  \BibitemOpen
  \bibfield  {author} {\bibinfo {author} {\bibfnamefont {S.}~\bibnamefont
  {Inouye}}, \bibinfo {author} {\bibfnamefont {M.~R.}\ \bibnamefont {Andrews}},
  \bibinfo {author} {\bibfnamefont {J.}~\bibnamefont {Stenger}}, \bibinfo
  {author} {\bibfnamefont {H.-J.}\ \bibnamefont {Miesner}}, \bibinfo {author}
  {\bibfnamefont {D.~M.}\ \bibnamefont {Stamper-Kurn}}, \ and\ \bibinfo
  {author} {\bibfnamefont {W.}~\bibnamefont {Ketterle}},\ }\href {\doibase
  10.1038/32354} {\bibfield  {journal} {\bibinfo  {journal} {Nature}\ }\textbf
  {\bibinfo {volume} {392}},\ \bibinfo {pages} {151} (\bibinfo {year}
  {1998})}\BibitemShut {NoStop}%
\bibitem [{\citenamefont {Chin}\ \emph {et~al.}(2010)\citenamefont {Chin},
  \citenamefont {Grimm}, \citenamefont {Julienne},\ and\ \citenamefont
  {Tiesinga}}]{Chin2010}%
  \BibitemOpen
  \bibfield  {author} {\bibinfo {author} {\bibfnamefont {C.}~\bibnamefont
  {Chin}}, \bibinfo {author} {\bibfnamefont {R.}~\bibnamefont {Grimm}},
  \bibinfo {author} {\bibfnamefont {P.}~\bibnamefont {Julienne}}, \ and\
  \bibinfo {author} {\bibfnamefont {E.}~\bibnamefont {Tiesinga}},\ }\href
  {\doibase 10.1103/revmodphys.82.1225} {\bibfield  {journal} {\bibinfo
  {journal} {Reviews of Modern Physics}\ }\textbf {\bibinfo {volume} {82}},\
  \bibinfo {pages} {1225} (\bibinfo {year} {2010})}\BibitemShut {NoStop}%
\bibitem [{\citenamefont {Regal}\ \emph {et~al.}(2003)\citenamefont {Regal},
  \citenamefont {Ticknor}, \citenamefont {Bohn},\ and\ \citenamefont
  {Jin}}]{Regal2003}%
  \BibitemOpen
  \bibfield  {author} {\bibinfo {author} {\bibfnamefont {C.~A.}\ \bibnamefont
  {Regal}}, \bibinfo {author} {\bibfnamefont {C.}~\bibnamefont {Ticknor}},
  \bibinfo {author} {\bibfnamefont {J.~L.}\ \bibnamefont {Bohn}}, \ and\
  \bibinfo {author} {\bibfnamefont {D.~S.}\ \bibnamefont {Jin}},\ }\href
  {\doibase 10.1038/nature01738} {\bibfield  {journal} {\bibinfo  {journal}
  {Nature}\ }\textbf {\bibinfo {volume} {424}},\ \bibinfo {pages} {47}
  (\bibinfo {year} {2003})}\BibitemShut {NoStop}%
\bibitem [{\citenamefont {Cubizolles}\ \emph {et~al.}(2003)\citenamefont
  {Cubizolles}, \citenamefont {Bourdel}, \citenamefont {Kokkelmans},
  \citenamefont {Shlyapnikov},\ and\ \citenamefont
  {Salomon}}]{Cubizolles03PRL}%
  \BibitemOpen
  \bibfield  {author} {\bibinfo {author} {\bibfnamefont {J.}~\bibnamefont
  {Cubizolles}}, \bibinfo {author} {\bibfnamefont {T.}~\bibnamefont {Bourdel}},
  \bibinfo {author} {\bibfnamefont {S.~J. J. M.~F.}\ \bibnamefont
  {Kokkelmans}}, \bibinfo {author} {\bibfnamefont {G.~V.}\ \bibnamefont
  {Shlyapnikov}}, \ and\ \bibinfo {author} {\bibfnamefont {C.}~\bibnamefont
  {Salomon}},\ }\href {\doibase 10.1103/PhysRevLett.91.240401} {\bibfield
  {journal} {\bibinfo  {journal} {Phys. Rev. Lett.}\ }\textbf {\bibinfo
  {volume} {91}},\ \bibinfo {pages} {240401} (\bibinfo {year}
  {2003})}\BibitemShut {NoStop}%
\bibitem [{\citenamefont {Jochim}\ \emph {et~al.}(2003)\citenamefont {Jochim},
  \citenamefont {Bartenstein}, \citenamefont {Altmeyer}, \citenamefont {Hendl},
  \citenamefont {Riedl}, \citenamefont {Chin}, \citenamefont {Denschlag},\ and\
  \citenamefont {Grimm}}]{Jochim03Science}%
  \BibitemOpen
  \bibfield  {author} {\bibinfo {author} {\bibfnamefont {S.}~\bibnamefont
  {Jochim}}, \bibinfo {author} {\bibfnamefont {M.}~\bibnamefont {Bartenstein}},
  \bibinfo {author} {\bibfnamefont {A.}~\bibnamefont {Altmeyer}}, \bibinfo
  {author} {\bibfnamefont {G.}~\bibnamefont {Hendl}}, \bibinfo {author}
  {\bibfnamefont {S.}~\bibnamefont {Riedl}}, \bibinfo {author} {\bibfnamefont
  {C.}~\bibnamefont {Chin}}, \bibinfo {author} {\bibfnamefont {J.~H.}\
  \bibnamefont {Denschlag}}, \ and\ \bibinfo {author} {\bibfnamefont
  {R.}~\bibnamefont {Grimm}},\ }\href {\doibase 10.1126/science.1093280}
  {\bibfield  {journal} {\bibinfo  {journal} {Science}\ }\textbf {\bibinfo
  {volume} {302}},\ \bibinfo {pages} {2101} (\bibinfo {year}
  {2003})}\BibitemShut {NoStop}%
\bibitem [{\citenamefont {Zwierlein}\ \emph {et~al.}(2003)\citenamefont
  {Zwierlein}, \citenamefont {Stan}, \citenamefont {Schunck}, \citenamefont
  {Raupach}, \citenamefont {Gupta}, \citenamefont {Hadzibabic},\ and\
  \citenamefont {Ketterle}}]{Zwierlein03PRL}%
  \BibitemOpen
  \bibfield  {author} {\bibinfo {author} {\bibfnamefont {M.~W.}\ \bibnamefont
  {Zwierlein}}, \bibinfo {author} {\bibfnamefont {C.~A.}\ \bibnamefont {Stan}},
  \bibinfo {author} {\bibfnamefont {C.~H.}\ \bibnamefont {Schunck}}, \bibinfo
  {author} {\bibfnamefont {S.~M.~F.}\ \bibnamefont {Raupach}}, \bibinfo
  {author} {\bibfnamefont {S.}~\bibnamefont {Gupta}}, \bibinfo {author}
  {\bibfnamefont {Z.}~\bibnamefont {Hadzibabic}}, \ and\ \bibinfo {author}
  {\bibfnamefont {W.}~\bibnamefont {Ketterle}},\ }\href {\doibase
  10.1103/PhysRevLett.91.250401} {\bibfield  {journal} {\bibinfo  {journal}
  {Phys. Rev. Lett.}\ }\textbf {\bibinfo {volume} {91}},\ \bibinfo {pages}
  {250401} (\bibinfo {year} {2003})}\BibitemShut {NoStop}%
\bibitem [{\citenamefont {Greiner}\ \emph {et~al.}(2003)\citenamefont
  {Greiner}, \citenamefont {Regal},\ and\ \citenamefont {Jin}}]{Greiner2003}%
  \BibitemOpen
  \bibfield  {author} {\bibinfo {author} {\bibfnamefont {M.}~\bibnamefont
  {Greiner}}, \bibinfo {author} {\bibfnamefont {C.~A.}\ \bibnamefont {Regal}},
  \ and\ \bibinfo {author} {\bibfnamefont {D.~S.}\ \bibnamefont {Jin}},\ }\href
  {\doibase 10.1038/nature02199} {\bibfield  {journal} {\bibinfo  {journal}
  {Nature}\ }\textbf {\bibinfo {volume} {426}},\ \bibinfo {pages} {537}
  (\bibinfo {year} {2003})}\BibitemShut {NoStop}%
\bibitem [{\citenamefont {Roati}\ \emph {et~al.}(2007)\citenamefont {Roati},
  \citenamefont {Zaccanti}, \citenamefont {D'Errico}, \citenamefont {Catani},
  \citenamefont {Modugno}, \citenamefont {Simoni}, \citenamefont {Inguscio},\
  and\ \citenamefont {Modugno}}]{Roati2007PRL}%
  \BibitemOpen
  \bibfield  {author} {\bibinfo {author} {\bibfnamefont {G.}~\bibnamefont
  {Roati}}, \bibinfo {author} {\bibfnamefont {M.}~\bibnamefont {Zaccanti}},
  \bibinfo {author} {\bibfnamefont {C.}~\bibnamefont {D'Errico}}, \bibinfo
  {author} {\bibfnamefont {J.}~\bibnamefont {Catani}}, \bibinfo {author}
  {\bibfnamefont {M.}~\bibnamefont {Modugno}}, \bibinfo {author} {\bibfnamefont
  {A.}~\bibnamefont {Simoni}}, \bibinfo {author} {\bibfnamefont
  {M.}~\bibnamefont {Inguscio}}, \ and\ \bibinfo {author} {\bibfnamefont
  {G.}~\bibnamefont {Modugno}},\ }\href {\doibase
  10.1103/PhysRevLett.99.010403} {\bibfield  {journal} {\bibinfo  {journal}
  {Phys. Rev. Lett.}\ }\textbf {\bibinfo {volume} {99}},\ \bibinfo {pages}
  {010403} (\bibinfo {year} {2007})}\BibitemShut {NoStop}%
\bibitem [{\citenamefont {Campbell}\ \emph {et~al.}(2010)\citenamefont
  {Campbell}, \citenamefont {Smith}, \citenamefont {Tammuz}, \citenamefont
  {Beattie}, \citenamefont {Moulder},\ and\ \citenamefont
  {Hadzibabic}}]{Campbell2010PRA}%
  \BibitemOpen
  \bibfield  {author} {\bibinfo {author} {\bibfnamefont {R.~L.~D.}\
  \bibnamefont {Campbell}}, \bibinfo {author} {\bibfnamefont {R.~P.}\
  \bibnamefont {Smith}}, \bibinfo {author} {\bibfnamefont {N.}~\bibnamefont
  {Tammuz}}, \bibinfo {author} {\bibfnamefont {S.}~\bibnamefont {Beattie}},
  \bibinfo {author} {\bibfnamefont {S.}~\bibnamefont {Moulder}}, \ and\
  \bibinfo {author} {\bibfnamefont {Z.}~\bibnamefont {Hadzibabic}},\ }\href
  {\doibase 10.1103/PhysRevA.82.063611} {\bibfield  {journal} {\bibinfo
  {journal} {Phys. Rev. A}\ }\textbf {\bibinfo {volume} {82}},\ \bibinfo
  {pages} {063611} (\bibinfo {year} {2010})}\BibitemShut {NoStop}%
\bibitem [{\citenamefont {Voges}\ \emph {et~al.}(2020)\citenamefont {Voges},
  \citenamefont {Gersema}, \citenamefont {Meyer~zum Alten~Borgloh},
  \citenamefont {Schulze}, \citenamefont {Hartmann}, \citenamefont {Zenesini},\
  and\ \citenamefont {Ospelkaus}}]{Voges20PRL}%
  \BibitemOpen
  \bibfield  {author} {\bibinfo {author} {\bibfnamefont {K.~K.}\ \bibnamefont
  {Voges}}, \bibinfo {author} {\bibfnamefont {P.}~\bibnamefont {Gersema}},
  \bibinfo {author} {\bibfnamefont {M.}~\bibnamefont {Meyer~zum
  Alten~Borgloh}}, \bibinfo {author} {\bibfnamefont {T.~A.}\ \bibnamefont
  {Schulze}}, \bibinfo {author} {\bibfnamefont {T.}~\bibnamefont {Hartmann}},
  \bibinfo {author} {\bibfnamefont {A.}~\bibnamefont {Zenesini}}, \ and\
  \bibinfo {author} {\bibfnamefont {S.}~\bibnamefont {Ospelkaus}},\ }\href
  {\doibase 10.1103/PhysRevLett.125.083401} {\bibfield  {journal} {\bibinfo
  {journal} {Phys. Rev. Lett.}\ }\textbf {\bibinfo {volume} {125}},\ \bibinfo
  {pages} {083401} (\bibinfo {year} {2020})}\BibitemShut {NoStop}%
\bibitem [{\citenamefont {Zhang}\ \emph {et~al.}(2021)\citenamefont {Zhang},
  \citenamefont {Chen}, \citenamefont {Yao},\ and\ \citenamefont
  {Chin}}]{Zhang2021}%
  \BibitemOpen
  \bibfield  {author} {\bibinfo {author} {\bibfnamefont {Z.}~\bibnamefont
  {Zhang}}, \bibinfo {author} {\bibfnamefont {L.}~\bibnamefont {Chen}},
  \bibinfo {author} {\bibfnamefont {K.-X.}\ \bibnamefont {Yao}}, \ and\
  \bibinfo {author} {\bibfnamefont {C.}~\bibnamefont {Chin}},\ }\href {\doibase
  10.1038/s41586-021-03443-0} {\bibfield  {journal} {\bibinfo  {journal}
  {Nature}\ }\textbf {\bibinfo {volume} {592}},\ \bibinfo {pages} {708}
  (\bibinfo {year} {2021})}\BibitemShut {NoStop}%
\bibitem [{\citenamefont {Eigen}\ \emph {et~al.}(2018)\citenamefont {Eigen},
  \citenamefont {Glidden}, \citenamefont {Lopes}, \citenamefont {Cornell},
  \citenamefont {Smith},\ and\ \citenamefont {Hadzibabic}}]{Eigen2018}%
  \BibitemOpen
  \bibfield  {author} {\bibinfo {author} {\bibfnamefont {C.}~\bibnamefont
  {Eigen}}, \bibinfo {author} {\bibfnamefont {J.~A.~P.}\ \bibnamefont
  {Glidden}}, \bibinfo {author} {\bibfnamefont {R.}~\bibnamefont {Lopes}},
  \bibinfo {author} {\bibfnamefont {E.~A.}\ \bibnamefont {Cornell}}, \bibinfo
  {author} {\bibfnamefont {R.~P.}\ \bibnamefont {Smith}}, \ and\ \bibinfo
  {author} {\bibfnamefont {Z.}~\bibnamefont {Hadzibabic}},\ }\href {\doibase
  10.1038/s41586-018-0674-1} {\bibfield  {journal} {\bibinfo  {journal}
  {Nature}\ }\textbf {\bibinfo {volume} {563}},\ \bibinfo {pages} {221}
  (\bibinfo {year} {2018})}\BibitemShut {NoStop}%
\bibitem [{\citenamefont {Mathey}\ \emph {et~al.}(2009)\citenamefont {Mathey},
  \citenamefont {Tiesinga}, \citenamefont {Julienne},\ and\ \citenamefont
  {Clark}}]{Mathey09PRA}%
  \BibitemOpen
  \bibfield  {author} {\bibinfo {author} {\bibfnamefont {L.}~\bibnamefont
  {Mathey}}, \bibinfo {author} {\bibfnamefont {E.}~\bibnamefont {Tiesinga}},
  \bibinfo {author} {\bibfnamefont {P.~S.}\ \bibnamefont {Julienne}}, \ and\
  \bibinfo {author} {\bibfnamefont {C.~W.}\ \bibnamefont {Clark}},\ }\href
  {\doibase 10.1103/PhysRevA.80.030702} {\bibfield  {journal} {\bibinfo
  {journal} {Phys. Rev. A}\ }\textbf {\bibinfo {volume} {80}},\ \bibinfo
  {pages} {030702} (\bibinfo {year} {2009})}\BibitemShut {NoStop}%
\bibitem [{\citenamefont {Peng}\ \emph {et~al.}(2021)\citenamefont {Peng},
  \citenamefont {Liu}, \citenamefont {Li},\ and\ \citenamefont
  {Luo}}]{Peng21arXiv}%
  \BibitemOpen
  \bibfield  {author} {\bibinfo {author} {\bibfnamefont {S.}~\bibnamefont
  {Peng}}, \bibinfo {author} {\bibfnamefont {H.}~\bibnamefont {Liu}}, \bibinfo
  {author} {\bibfnamefont {J.}~\bibnamefont {Li}}, \ and\ \bibinfo {author}
  {\bibfnamefont {L.}~\bibnamefont {Luo}},\ }\href
  {https://arxiv.org/abs/2107.07078} {\enquote {\bibinfo {title} {Cooling a
  fermi gas with three-body recombination near a narrow feshbach resonance},}\
  } (\bibinfo {year} {2021}),\ \Eprint {http://arxiv.org/abs/2107.07078}
  {arXiv:2107.07078 [cond-mat.quant-gas]} \BibitemShut {NoStop}%
\bibitem [{\citenamefont {D{\textquotesingle}Errico}\ \emph
  {et~al.}(2007)\citenamefont {D{\textquotesingle}Errico}, \citenamefont
  {Zaccanti}, \citenamefont {Fattori}, \citenamefont {Roati}, \citenamefont
  {Inguscio}, \citenamefont {Modugno},\ and\ \citenamefont
  {Simoni}}]{DErrico07NJP}%
  \BibitemOpen
  \bibfield  {author} {\bibinfo {author} {\bibfnamefont {C.}~\bibnamefont
  {D{\textquotesingle}Errico}}, \bibinfo {author} {\bibfnamefont
  {M.}~\bibnamefont {Zaccanti}}, \bibinfo {author} {\bibfnamefont
  {M.}~\bibnamefont {Fattori}}, \bibinfo {author} {\bibfnamefont
  {G.}~\bibnamefont {Roati}}, \bibinfo {author} {\bibfnamefont
  {M.}~\bibnamefont {Inguscio}}, \bibinfo {author} {\bibfnamefont
  {G.}~\bibnamefont {Modugno}}, \ and\ \bibinfo {author} {\bibfnamefont
  {A.}~\bibnamefont {Simoni}},\ }\href {\doibase 10.1088/1367-2630/9/7/223}
  {\bibfield  {journal} {\bibinfo  {journal} {New Journal of Physics}\ }\textbf
  {\bibinfo {volume} {9}},\ \bibinfo {pages} {223} (\bibinfo {year}
  {2007})}\BibitemShut {NoStop}%
\bibitem [{\citenamefont {Salomon}\ \emph {et~al.}(2014)\citenamefont
  {Salomon}, \citenamefont {Fouch{\'{e}}}, \citenamefont {Lepoutre},
  \citenamefont {Aspect},\ and\ \citenamefont {Bourdel}}]{Salomon14PRA}%
  \BibitemOpen
  \bibfield  {author} {\bibinfo {author} {\bibfnamefont {G.}~\bibnamefont
  {Salomon}}, \bibinfo {author} {\bibfnamefont {L.}~\bibnamefont
  {Fouch{\'{e}}}}, \bibinfo {author} {\bibfnamefont {S.}~\bibnamefont
  {Lepoutre}}, \bibinfo {author} {\bibfnamefont {A.}~\bibnamefont {Aspect}}, \
  and\ \bibinfo {author} {\bibfnamefont {T.}~\bibnamefont {Bourdel}},\ }\href
  {\doibase 10.1103/physreva.90.033405} {\bibfield  {journal} {\bibinfo
  {journal} {Physical Review A}\ }\textbf {\bibinfo {volume} {90}} (\bibinfo
  {year} {2014}),\ 10.1103/physreva.90.033405}\BibitemShut {NoStop}%
\bibitem [{\citenamefont {Schlippert}\ \emph {et~al.}(2014)\citenamefont
  {Schlippert}, \citenamefont {Hartwig}, \citenamefont {Albers}, \citenamefont
  {Richardson}, \citenamefont {Schubert}, \citenamefont {Roura}, \citenamefont
  {Schleich}, \citenamefont {Ertmer},\ and\ \citenamefont
  {Rasel}}]{Schlippert14PRL}%
  \BibitemOpen
  \bibfield  {author} {\bibinfo {author} {\bibfnamefont {D.}~\bibnamefont
  {Schlippert}}, \bibinfo {author} {\bibfnamefont {J.}~\bibnamefont {Hartwig}},
  \bibinfo {author} {\bibfnamefont {H.}~\bibnamefont {Albers}}, \bibinfo
  {author} {\bibfnamefont {L.}~\bibnamefont {Richardson}}, \bibinfo {author}
  {\bibfnamefont {C.}~\bibnamefont {Schubert}}, \bibinfo {author}
  {\bibfnamefont {A.}~\bibnamefont {Roura}}, \bibinfo {author} {\bibfnamefont
  {W.}~\bibnamefont {Schleich}}, \bibinfo {author} {\bibfnamefont
  {W.}~\bibnamefont {Ertmer}}, \ and\ \bibinfo {author} {\bibfnamefont
  {E.}~\bibnamefont {Rasel}},\ }\href@noop {} {\bibfield  {journal} {\bibinfo
  {journal} {Phys. Rev. Lett.}\ }\textbf {\bibinfo {volume} {112}},\ \bibinfo
  {pages} {203002} (\bibinfo {year} {2014})}\BibitemShut {NoStop}%
\bibitem [{\citenamefont {Albers}\ \emph {et~al.}(2020)\citenamefont {Albers},
  \citenamefont {Herbst}, \citenamefont {Richardson}, \citenamefont {Heine},
  \citenamefont {Nath}, \citenamefont {Hartwig}, \citenamefont {Schubert},
  \citenamefont {Vogt}, \citenamefont {Woltmann}, \citenamefont {Lämmerzahl},
  \citenamefont {Herrmann}, \citenamefont {Ertmer}, \citenamefont {Rasel},\
  and\ \citenamefont {Schlippert}}]{Albers2020EPJD}%
  \BibitemOpen
  \bibfield  {author} {\bibinfo {author} {\bibfnamefont {H.}~\bibnamefont
  {Albers}}, \bibinfo {author} {\bibfnamefont {A.}~\bibnamefont {Herbst}},
  \bibinfo {author} {\bibfnamefont {L.~L.}\ \bibnamefont {Richardson}},
  \bibinfo {author} {\bibfnamefont {H.}~\bibnamefont {Heine}}, \bibinfo
  {author} {\bibfnamefont {D.}~\bibnamefont {Nath}}, \bibinfo {author}
  {\bibfnamefont {J.}~\bibnamefont {Hartwig}}, \bibinfo {author} {\bibfnamefont
  {C.}~\bibnamefont {Schubert}}, \bibinfo {author} {\bibfnamefont
  {C.}~\bibnamefont {Vogt}}, \bibinfo {author} {\bibfnamefont {M.}~\bibnamefont
  {Woltmann}}, \bibinfo {author} {\bibfnamefont {C.}~\bibnamefont
  {Lämmerzahl}}, \bibinfo {author} {\bibfnamefont {S.}~\bibnamefont
  {Herrmann}}, \bibinfo {author} {\bibfnamefont {W.}~\bibnamefont {Ertmer}},
  \bibinfo {author} {\bibfnamefont {E.~M.}\ \bibnamefont {Rasel}}, \ and\
  \bibinfo {author} {\bibfnamefont {D.}~\bibnamefont {Schlippert}},\ }\href
  {\doibase 10.1140/epjd/e2020-10132-6} {\bibfield  {journal} {\bibinfo
  {journal} {The European Physical Journal D}\ }\textbf {\bibinfo {volume}
  {74}} (\bibinfo {year} {2020}),\ 10.1140/epjd/e2020-10132-6}\BibitemShut
  {NoStop}%
\bibitem [{\citenamefont {Baillard}\ \emph {et~al.}(2006)\citenamefont
  {Baillard}, \citenamefont {Gauguet}, \citenamefont {Bize}, \citenamefont
  {Lemonde}, \citenamefont {Laurent}, \citenamefont {Clairon},\ and\
  \citenamefont {Rosenbusch}}]{Baillard06OC}%
  \BibitemOpen
  \bibfield  {author} {\bibinfo {author} {\bibfnamefont {X.}~\bibnamefont
  {Baillard}}, \bibinfo {author} {\bibfnamefont {A.}~\bibnamefont {Gauguet}},
  \bibinfo {author} {\bibfnamefont {S.}~\bibnamefont {Bize}}, \bibinfo {author}
  {\bibfnamefont {P.}~\bibnamefont {Lemonde}}, \bibinfo {author} {\bibfnamefont
  {P.}~\bibnamefont {Laurent}}, \bibinfo {author} {\bibfnamefont
  {A.}~\bibnamefont {Clairon}}, \ and\ \bibinfo {author} {\bibfnamefont
  {P.}~\bibnamefont {Rosenbusch}},\ }\href {\doibase
  10.1016/j.optcom.2006.05.011} {\bibfield  {journal} {\bibinfo  {journal}
  {Opt. Comm.}\ }\textbf {\bibinfo {volume} {266}},\ \bibinfo {pages} {609}
  (\bibinfo {year} {2006})}\BibitemShut {NoStop}%
\bibitem [{\citenamefont {Gilowski}\ \emph {et~al.}(2007)\citenamefont
  {Gilowski}, \citenamefont {Schubert}, \citenamefont {Zaiser}, \citenamefont
  {Herr}, \citenamefont {W\"ubbena}, \citenamefont {{T. Wendrich}},
  \citenamefont {{T. M\"uller}}, \citenamefont {Rasel},\ and\ \citenamefont
  {Ertmer}}]{Gilowski07OC}%
  \BibitemOpen
  \bibfield  {author} {\bibinfo {author} {\bibfnamefont {M.}~\bibnamefont
  {Gilowski}}, \bibinfo {author} {\bibfnamefont {C.}~\bibnamefont {Schubert}},
  \bibinfo {author} {\bibfnamefont {M.}~\bibnamefont {Zaiser}}, \bibinfo
  {author} {\bibfnamefont {W.}~\bibnamefont {Herr}}, \bibinfo {author}
  {\bibfnamefont {T.}~\bibnamefont {W\"ubbena}}, \bibinfo {author}
  {\bibnamefont {{T. Wendrich}}}, \bibinfo {author} {\bibnamefont {{T.
  M\"uller}}}, \bibinfo {author} {\bibfnamefont {E.}~\bibnamefont {Rasel}}, \
  and\ \bibinfo {author} {\bibfnamefont {W.}~\bibnamefont {Ertmer}},\ }\href
  {\doibase 10.1016/j.optcom.2007.08.043} {\bibfield  {journal} {\bibinfo
  {journal} {Opt. Comm.}\ }\textbf {\bibinfo {volume} {280}},\ \bibinfo {pages}
  {443} (\bibinfo {year} {2007})}\BibitemShut {NoStop}%
\bibitem [{\citenamefont {McCarron}\ \emph {et~al.}(2008)\citenamefont
  {McCarron}, \citenamefont {King},\ and\ \citenamefont
  {Cornish}}]{McCarron2008}%
  \BibitemOpen
  \bibfield  {author} {\bibinfo {author} {\bibfnamefont {D.~J.}\ \bibnamefont
  {McCarron}}, \bibinfo {author} {\bibfnamefont {S.~A.}\ \bibnamefont {King}},
  \ and\ \bibinfo {author} {\bibfnamefont {S.~L.}\ \bibnamefont {Cornish}},\
  }\href {\doibase 10.1088/0957-0233/19/10/105601} {\bibfield  {journal}
  {\bibinfo  {journal} {Measurement Science and Technology}\ }\textbf {\bibinfo
  {volume} {19}},\ \bibinfo {pages} {105601} (\bibinfo {year}
  {2008})}\BibitemShut {NoStop}%
\bibitem [{\citenamefont {Albers}(2020)}]{Albers2020phd}%
  \BibitemOpen
  \bibfield  {author} {\bibinfo {author} {\bibfnamefont {H.}~\bibnamefont
  {Albers}},\ }\emph {\bibinfo {title} {Time-averaged optical potentials for
  creating and shaping Bose-Einstein condensates}},\ \href {\doibase
  10.15488/10073} {Ph.D. thesis},\ \bibinfo  {school} {Leibniz Universit\"at
  Hannover} (\bibinfo {year} {2020})\BibitemShut {NoStop}%
\bibitem [{\citenamefont {Zaiser}(2010)}]{Zaiser2010PHD}%
  \BibitemOpen
  \bibfield  {author} {\bibinfo {author} {\bibfnamefont {M.}~\bibnamefont
  {Zaiser}},\ }\emph {\bibinfo {title} {Eine Quelle quantenentarteter Gase für
  die Atominterferometrie}},\ \href {\doibase 10.15488/7563} {\bibinfo {type}
  {Dissertation}},\ \bibinfo  {school} {Leibniz Universit\"at Hannover}
  (\bibinfo {year} {2010})\BibitemShut {NoStop}%
\bibitem [{\citenamefont {Weber}\ \emph {et~al.}(2003)\citenamefont {Weber},
  \citenamefont {Herbig}, \citenamefont {Mark}, \citenamefont {N\"agerl},\ and\
  \citenamefont {Grimm}}]{Weber03PRL}%
  \BibitemOpen
  \bibfield  {author} {\bibinfo {author} {\bibfnamefont {T.}~\bibnamefont
  {Weber}}, \bibinfo {author} {\bibfnamefont {J.}~\bibnamefont {Herbig}},
  \bibinfo {author} {\bibfnamefont {M.}~\bibnamefont {Mark}}, \bibinfo {author}
  {\bibfnamefont {H.-C.}\ \bibnamefont {N\"agerl}}, \ and\ \bibinfo {author}
  {\bibfnamefont {R.}~\bibnamefont {Grimm}},\ }\href {\doibase
  10.1103/PhysRevLett.91.123201} {\bibfield  {journal} {\bibinfo  {journal}
  {Phys. Rev. Lett.}\ }\textbf {\bibinfo {volume} {91}},\ \bibinfo {pages}
  {123201} (\bibinfo {year} {2003})}\BibitemShut {NoStop}%
\bibitem [{\citenamefont {Fedichev}\ \emph {et~al.}(1996)\citenamefont
  {Fedichev}, \citenamefont {Reynolds},\ and\ \citenamefont
  {Shlyapnikov}}]{Fedichev1996PRL}%
  \BibitemOpen
  \bibfield  {author} {\bibinfo {author} {\bibfnamefont {P.~O.}\ \bibnamefont
  {Fedichev}}, \bibinfo {author} {\bibfnamefont {M.~W.}\ \bibnamefont
  {Reynolds}}, \ and\ \bibinfo {author} {\bibfnamefont {G.~V.}\ \bibnamefont
  {Shlyapnikov}},\ }\href {\doibase 10.1103/PhysRevLett.77.2921} {\bibfield
  {journal} {\bibinfo  {journal} {Phys. Rev. Lett.}\ }\textbf {\bibinfo
  {volume} {77}},\ \bibinfo {pages} {2921} (\bibinfo {year}
  {1996})}\BibitemShut {NoStop}%
\bibitem [{\citenamefont {Tiemann}\ \emph {et~al.}(2020)\citenamefont
  {Tiemann}, \citenamefont {Gersema}, \citenamefont {Voges}, \citenamefont
  {Hartmann}, \citenamefont {Zenesini},\ and\ \citenamefont
  {Ospelkaus}}]{Tiemann20PRR}%
  \BibitemOpen
  \bibfield  {author} {\bibinfo {author} {\bibfnamefont {E.}~\bibnamefont
  {Tiemann}}, \bibinfo {author} {\bibfnamefont {P.}~\bibnamefont {Gersema}},
  \bibinfo {author} {\bibfnamefont {K.~K.}\ \bibnamefont {Voges}}, \bibinfo
  {author} {\bibfnamefont {T.}~\bibnamefont {Hartmann}}, \bibinfo {author}
  {\bibfnamefont {A.}~\bibnamefont {Zenesini}}, \ and\ \bibinfo {author}
  {\bibfnamefont {S.}~\bibnamefont {Ospelkaus}},\ }\href {\doibase
  10.1103/PhysRevResearch.2.013366} {\bibfield  {journal} {\bibinfo  {journal}
  {Phys. Rev. Research}\ }\textbf {\bibinfo {volume} {2}},\ \bibinfo {pages}
  {013366} (\bibinfo {year} {2020})}\BibitemShut {NoStop}%
\bibitem [{\citenamefont {Salomon}\ \emph {et~al.}(2013)\citenamefont
  {Salomon}, \citenamefont {Fouch{\'e}}, \citenamefont {Wang}, \citenamefont
  {Aspect}, \citenamefont {Bouyer},\ and\ \citenamefont
  {Bourdel}}]{Salomon13EPL}%
  \BibitemOpen
  \bibfield  {author} {\bibinfo {author} {\bibfnamefont {G.}~\bibnamefont
  {Salomon}}, \bibinfo {author} {\bibfnamefont {L.}~\bibnamefont {Fouch{\'e}}},
  \bibinfo {author} {\bibfnamefont {P.}~\bibnamefont {Wang}}, \bibinfo {author}
  {\bibfnamefont {A.}~\bibnamefont {Aspect}}, \bibinfo {author} {\bibfnamefont
  {P.}~\bibnamefont {Bouyer}}, \ and\ \bibinfo {author} {\bibfnamefont
  {T.}~\bibnamefont {Bourdel}},\ }\href
  {http://stacks.iop.org/0295-5075/104/i=6/a=63002} {\bibfield  {journal}
  {\bibinfo  {journal} {Eur. Phys. Lett.}\ }\textbf {\bibinfo {volume} {104}},\
  \bibinfo {pages} {63002} (\bibinfo {year} {2013})}\BibitemShut {NoStop}%
\bibitem [{\citenamefont {Gaaloul}\ \emph {et~al.}(2006)\citenamefont
  {Gaaloul}, \citenamefont {Suzor-Weiner}, \citenamefont {Pruvost},
  \citenamefont {Telmini},\ and\ \citenamefont {Charron}}]{Gaaloul2006PRA}%
  \BibitemOpen
  \bibfield  {author} {\bibinfo {author} {\bibfnamefont {N.}~\bibnamefont
  {Gaaloul}}, \bibinfo {author} {\bibfnamefont {A.}~\bibnamefont
  {Suzor-Weiner}}, \bibinfo {author} {\bibfnamefont {L.}~\bibnamefont
  {Pruvost}}, \bibinfo {author} {\bibfnamefont {M.}~\bibnamefont {Telmini}}, \
  and\ \bibinfo {author} {\bibfnamefont {E.}~\bibnamefont {Charron}},\ }\href
  {\doibase 10.1103/PhysRevA.74.023620} {\bibfield  {journal} {\bibinfo
  {journal} {Phys. Rev. A}\ }\textbf {\bibinfo {volume} {74}},\ \bibinfo
  {pages} {023620} (\bibinfo {year} {2006})}\BibitemShut {NoStop}%
\bibitem [{\citenamefont {Antoni-Micollier}\ \emph {et~al.}(2017)\citenamefont
  {Antoni-Micollier}, \citenamefont {Barrett}, \citenamefont {Chichet},
  \citenamefont {Condon}, \citenamefont {Battelier}, \citenamefont
  {Landragin},\ and\ \citenamefont {Bouyer}}]{Antoni-Micollier2017}%
  \BibitemOpen
  \bibfield  {author} {\bibinfo {author} {\bibfnamefont {L.}~\bibnamefont
  {Antoni-Micollier}}, \bibinfo {author} {\bibfnamefont {B.}~\bibnamefont
  {Barrett}}, \bibinfo {author} {\bibfnamefont {L.}~\bibnamefont {Chichet}},
  \bibinfo {author} {\bibfnamefont {G.}~\bibnamefont {Condon}}, \bibinfo
  {author} {\bibfnamefont {B.}~\bibnamefont {Battelier}}, \bibinfo {author}
  {\bibfnamefont {A.}~\bibnamefont {Landragin}}, \ and\ \bibinfo {author}
  {\bibfnamefont {P.}~\bibnamefont {Bouyer}},\ }\href {\doibase
  10.1103/physreva.96.023608} {\bibfield  {journal} {\bibinfo  {journal}
  {Physical Review A}\ }\textbf {\bibinfo {volume} {96}} (\bibinfo {year}
  {2017}),\ 10.1103/physreva.96.023608}\BibitemShut {NoStop}%
\bibitem [{\citenamefont {Tiesinga}\ \emph {et~al.}(1993)\citenamefont
  {Tiesinga}, \citenamefont {Verhaar},\ and\ \citenamefont
  {Stoof}}]{Tiesinga1993PRA}%
  \BibitemOpen
  \bibfield  {author} {\bibinfo {author} {\bibfnamefont {E.}~\bibnamefont
  {Tiesinga}}, \bibinfo {author} {\bibfnamefont {B.~J.}\ \bibnamefont
  {Verhaar}}, \ and\ \bibinfo {author} {\bibfnamefont {H.~T.~C.}\ \bibnamefont
  {Stoof}},\ }\href {\doibase 10.1103/PhysRevA.47.4114} {\bibfield  {journal}
  {\bibinfo  {journal} {Phys. Rev. A}\ }\textbf {\bibinfo {volume} {47}},\
  \bibinfo {pages} {4114} (\bibinfo {year} {1993})}\BibitemShut {NoStop}%
\bibitem [{\citenamefont {Ketterle}\ and\ \citenamefont
  {Druten}(1996)}]{Ketterle96AAMOP}%
  \BibitemOpen
  \bibfield  {author} {\bibinfo {author} {\bibfnamefont {W.}~\bibnamefont
  {Ketterle}}\ and\ \bibinfo {author} {\bibfnamefont {N.~V.}\ \bibnamefont
  {Druten}},\ }in\ \href {\doibase 10.1016/s1049-250x(08)60101-9} {\emph
  {\bibinfo {booktitle} {Advances In Atomic, Molecular, and Optical Physics}}}\
  (\bibinfo  {publisher} {Elsevier},\ \bibinfo {year} {1996})\ pp.\ \bibinfo
  {pages} {181--236}\BibitemShut {NoStop}%
\bibitem [{\citenamefont {Simonelli}\ \emph {et~al.}(2019)\citenamefont
  {Simonelli}, \citenamefont {Neri}, \citenamefont {Ciamei}, \citenamefont
  {Goti}, \citenamefont {Inguscio}, \citenamefont {Trenkwalder},\ and\
  \citenamefont {Zaccanti}}]{Simonelli2019OptExpress}%
  \BibitemOpen
  \bibfield  {author} {\bibinfo {author} {\bibfnamefont {C.}~\bibnamefont
  {Simonelli}}, \bibinfo {author} {\bibfnamefont {E.}~\bibnamefont {Neri}},
  \bibinfo {author} {\bibfnamefont {A.}~\bibnamefont {Ciamei}}, \bibinfo
  {author} {\bibfnamefont {I.}~\bibnamefont {Goti}}, \bibinfo {author}
  {\bibfnamefont {M.}~\bibnamefont {Inguscio}}, \bibinfo {author}
  {\bibfnamefont {A.}~\bibnamefont {Trenkwalder}}, \ and\ \bibinfo {author}
  {\bibfnamefont {M.}~\bibnamefont {Zaccanti}},\ }\href {\doibase
  10.1364/OE.27.027215} {\bibfield  {journal} {\bibinfo  {journal} {Opt.
  Express}\ }\textbf {\bibinfo {volume} {27}},\ \bibinfo {pages} {27215}
  (\bibinfo {year} {2019})}\BibitemShut {NoStop}%
\bibitem [{\citenamefont {Catani}\ \emph {et~al.}(2006)\citenamefont {Catani},
  \citenamefont {Maioli}, \citenamefont {De~Sarlo}, \citenamefont {Minardi},\
  and\ \citenamefont {Inguscio}}]{Catani2006PRA}%
  \BibitemOpen
  \bibfield  {author} {\bibinfo {author} {\bibfnamefont {J.}~\bibnamefont
  {Catani}}, \bibinfo {author} {\bibfnamefont {P.}~\bibnamefont {Maioli}},
  \bibinfo {author} {\bibfnamefont {L.}~\bibnamefont {De~Sarlo}}, \bibinfo
  {author} {\bibfnamefont {F.}~\bibnamefont {Minardi}}, \ and\ \bibinfo
  {author} {\bibfnamefont {M.}~\bibnamefont {Inguscio}},\ }\href {\doibase
  10.1103/PhysRevA.73.033415} {\bibfield  {journal} {\bibinfo  {journal} {Phys.
  Rev. A}\ }\textbf {\bibinfo {volume} {73}},\ \bibinfo {pages} {033415}
  (\bibinfo {year} {2006})}\BibitemShut {NoStop}%
\bibitem [{\citenamefont {Lasner}\ \emph {et~al.}(2021)\citenamefont {Lasner},
  \citenamefont {Mitra}, \citenamefont {Hiradfar}, \citenamefont {Augenbraun},
  \citenamefont {Cheuk}, \citenamefont {Lee}, \citenamefont {Prabhu},\ and\
  \citenamefont {Doyle}}]{Doyle2021PRA}%
  \BibitemOpen
  \bibfield  {author} {\bibinfo {author} {\bibfnamefont {Z.}~\bibnamefont
  {Lasner}}, \bibinfo {author} {\bibfnamefont {D.}~\bibnamefont {Mitra}},
  \bibinfo {author} {\bibfnamefont {M.}~\bibnamefont {Hiradfar}}, \bibinfo
  {author} {\bibfnamefont {B.}~\bibnamefont {Augenbraun}}, \bibinfo {author}
  {\bibfnamefont {L.}~\bibnamefont {Cheuk}}, \bibinfo {author} {\bibfnamefont
  {E.}~\bibnamefont {Lee}}, \bibinfo {author} {\bibfnamefont {S.}~\bibnamefont
  {Prabhu}}, \ and\ \bibinfo {author} {\bibfnamefont {J.}~\bibnamefont
  {Doyle}},\ }\href {\doibase 10.1103/PhysRevA.104.063305} {\bibfield
  {journal} {\bibinfo  {journal} {Phys. Rev. A}\ }\textbf {\bibinfo {volume}
  {104}},\ \bibinfo {pages} {063305} (\bibinfo {year} {2021})}\BibitemShut
  {NoStop}%
\bibitem [{\citenamefont {Marte}\ \emph {et~al.}(2002)\citenamefont {Marte},
  \citenamefont {Volz}, \citenamefont {Schuster}, \citenamefont {D\"urr},
  \citenamefont {Rempe}, \citenamefont {van Kempen},\ and\ \citenamefont
  {Verhaar}}]{Marte2002PRL}%
  \BibitemOpen
  \bibfield  {author} {\bibinfo {author} {\bibfnamefont {A.}~\bibnamefont
  {Marte}}, \bibinfo {author} {\bibfnamefont {T.}~\bibnamefont {Volz}},
  \bibinfo {author} {\bibfnamefont {J.}~\bibnamefont {Schuster}}, \bibinfo
  {author} {\bibfnamefont {S.}~\bibnamefont {D\"urr}}, \bibinfo {author}
  {\bibfnamefont {G.}~\bibnamefont {Rempe}}, \bibinfo {author} {\bibfnamefont
  {E.~G.~M.}\ \bibnamefont {van Kempen}}, \ and\ \bibinfo {author}
  {\bibfnamefont {B.~J.}\ \bibnamefont {Verhaar}},\ }\href {\doibase
  10.1103/PhysRevLett.89.283202} {\bibfield  {journal} {\bibinfo  {journal}
  {Phys. Rev. Lett.}\ }\textbf {\bibinfo {volume} {89}},\ \bibinfo {pages}
  {283202} (\bibinfo {year} {2002})}\BibitemShut {NoStop}%
\bibitem [{\citenamefont {Roberts}\ \emph {et~al.}(1998)\citenamefont
  {Roberts}, \citenamefont {Claussen}, \citenamefont {Burke}, \citenamefont
  {Greene}, \citenamefont {Cornell},\ and\ \citenamefont
  {Wieman}}]{Roberts1998PRL}%
  \BibitemOpen
  \bibfield  {author} {\bibinfo {author} {\bibfnamefont {J.~L.}\ \bibnamefont
  {Roberts}}, \bibinfo {author} {\bibfnamefont {N.~R.}\ \bibnamefont
  {Claussen}}, \bibinfo {author} {\bibfnamefont {J.~P.}\ \bibnamefont {Burke}},
  \bibinfo {author} {\bibfnamefont {C.~H.}\ \bibnamefont {Greene}}, \bibinfo
  {author} {\bibfnamefont {E.~A.}\ \bibnamefont {Cornell}}, \ and\ \bibinfo
  {author} {\bibfnamefont {C.~E.}\ \bibnamefont {Wieman}},\ }\href {\doibase
  10.1103/PhysRevLett.81.5109} {\bibfield  {journal} {\bibinfo  {journal}
  {Phys. Rev. Lett.}\ }\textbf {\bibinfo {volume} {81}},\ \bibinfo {pages}
  {5109} (\bibinfo {year} {1998})}\BibitemShut {NoStop}%
\bibitem [{\citenamefont {Knoop}\ \emph {et~al.}(2011)\citenamefont {Knoop},
  \citenamefont {Schuster}, \citenamefont {Scelle}, \citenamefont {Trautmann},
  \citenamefont {Appmeier}, \citenamefont {Oberthaler}, \citenamefont
  {Tiesinga},\ and\ \citenamefont {Tiemann}}]{Knoop2011PRA}%
  \BibitemOpen
  \bibfield  {author} {\bibinfo {author} {\bibfnamefont {S.}~\bibnamefont
  {Knoop}}, \bibinfo {author} {\bibfnamefont {T.}~\bibnamefont {Schuster}},
  \bibinfo {author} {\bibfnamefont {R.}~\bibnamefont {Scelle}}, \bibinfo
  {author} {\bibfnamefont {A.}~\bibnamefont {Trautmann}}, \bibinfo {author}
  {\bibfnamefont {J.}~\bibnamefont {Appmeier}}, \bibinfo {author}
  {\bibfnamefont {M.~K.}\ \bibnamefont {Oberthaler}}, \bibinfo {author}
  {\bibfnamefont {E.}~\bibnamefont {Tiesinga}}, \ and\ \bibinfo {author}
  {\bibfnamefont {E.}~\bibnamefont {Tiemann}},\ }\href {\doibase
  10.1103/PhysRevA.83.042704} {\bibfield  {journal} {\bibinfo  {journal} {Phys.
  Rev. A}\ }\textbf {\bibinfo {volume} {83}},\ \bibinfo {pages} {042704}
  (\bibinfo {year} {2011})}\BibitemShut {NoStop}%
\bibitem [{\citenamefont {Semeghini}\ \emph {et~al.}(2018)\citenamefont
  {Semeghini}, \citenamefont {Ferioli}, \citenamefont {Masi}, \citenamefont
  {Mazzinghi}, \citenamefont {Wolswijk}, \citenamefont {Minardi}, \citenamefont
  {Modugno}, \citenamefont {Modugno}, \citenamefont {Inguscio},\ and\
  \citenamefont {Fattori}}]{Semeghini2018PRL}%
  \BibitemOpen
  \bibfield  {author} {\bibinfo {author} {\bibfnamefont {G.}~\bibnamefont
  {Semeghini}}, \bibinfo {author} {\bibfnamefont {G.}~\bibnamefont {Ferioli}},
  \bibinfo {author} {\bibfnamefont {L.}~\bibnamefont {Masi}}, \bibinfo {author}
  {\bibfnamefont {C.}~\bibnamefont {Mazzinghi}}, \bibinfo {author}
  {\bibfnamefont {L.}~\bibnamefont {Wolswijk}}, \bibinfo {author}
  {\bibfnamefont {F.}~\bibnamefont {Minardi}}, \bibinfo {author} {\bibfnamefont
  {M.}~\bibnamefont {Modugno}}, \bibinfo {author} {\bibfnamefont
  {G.}~\bibnamefont {Modugno}}, \bibinfo {author} {\bibfnamefont
  {M.}~\bibnamefont {Inguscio}}, \ and\ \bibinfo {author} {\bibfnamefont
  {M.}~\bibnamefont {Fattori}},\ }\href {\doibase
  10.1103/PhysRevLett.120.235301} {\bibfield  {journal} {\bibinfo  {journal}
  {Phys. Rev. Lett.}\ }\textbf {\bibinfo {volume} {120}},\ \bibinfo {pages}
  {235301} (\bibinfo {year} {2018})}\BibitemShut {NoStop}%
\bibitem [{\citenamefont {Cabrera}\ \emph {et~al.}(2018)\citenamefont
  {Cabrera}, \citenamefont {Tanzi}, \citenamefont {Sanz}, \citenamefont
  {Naylor}, \citenamefont {Thomas}, \citenamefont {Cheiney},\ and\
  \citenamefont {Tarruell}}]{Cabrera2018Science}%
  \BibitemOpen
  \bibfield  {author} {\bibinfo {author} {\bibfnamefont {C.~R.}\ \bibnamefont
  {Cabrera}}, \bibinfo {author} {\bibfnamefont {L.}~\bibnamefont {Tanzi}},
  \bibinfo {author} {\bibfnamefont {J.}~\bibnamefont {Sanz}}, \bibinfo {author}
  {\bibfnamefont {B.}~\bibnamefont {Naylor}}, \bibinfo {author} {\bibfnamefont
  {P.}~\bibnamefont {Thomas}}, \bibinfo {author} {\bibfnamefont
  {P.}~\bibnamefont {Cheiney}}, \ and\ \bibinfo {author} {\bibfnamefont
  {L.}~\bibnamefont {Tarruell}},\ }\href {\doibase 10.1126/science.aao5686}
  {\bibfield  {journal} {\bibinfo  {journal} {Science}\ }\textbf {\bibinfo
  {volume} {359}},\ \bibinfo {pages} {301} (\bibinfo {year}
  {2018})}\BibitemShut {NoStop}%
\bibitem [{\citenamefont {Hung}\ \emph {et~al.}(2013)\citenamefont {Hung},
  \citenamefont {Gurarie},\ and\ \citenamefont {Chin}}]{Hung2013Science}%
  \BibitemOpen
  \bibfield  {author} {\bibinfo {author} {\bibfnamefont {C.-L.}\ \bibnamefont
  {Hung}}, \bibinfo {author} {\bibfnamefont {V.}~\bibnamefont {Gurarie}}, \
  and\ \bibinfo {author} {\bibfnamefont {C.}~\bibnamefont {Chin}},\ }\href
  {\doibase 10.1126/science.1237557} {\bibfield  {journal} {\bibinfo  {journal}
  {Science}\ }\textbf {\bibinfo {volume} {341}},\ \bibinfo {pages} {1213}
  (\bibinfo {year} {2013})}\BibitemShut {NoStop}%
\bibitem [{\citenamefont {Hensel}\ \emph {et~al.}(2021)\citenamefont {Hensel},
  \citenamefont {Loriani}, \citenamefont {Schubert}, \citenamefont {Fitzek},
  \citenamefont {Abend}, \citenamefont {Ahlers}, \citenamefont {Siem{\ss}},
  \citenamefont {Hammerer}, \citenamefont {Rasel},\ and\ \citenamefont
  {Gaaloul}}]{Hensel2021}%
  \BibitemOpen
  \bibfield  {author} {\bibinfo {author} {\bibfnamefont {T.}~\bibnamefont
  {Hensel}}, \bibinfo {author} {\bibfnamefont {S.}~\bibnamefont {Loriani}},
  \bibinfo {author} {\bibfnamefont {C.}~\bibnamefont {Schubert}}, \bibinfo
  {author} {\bibfnamefont {F.}~\bibnamefont {Fitzek}}, \bibinfo {author}
  {\bibfnamefont {S.}~\bibnamefont {Abend}}, \bibinfo {author} {\bibfnamefont
  {H.}~\bibnamefont {Ahlers}}, \bibinfo {author} {\bibfnamefont {J.~N.}\
  \bibnamefont {Siem{\ss}}}, \bibinfo {author} {\bibfnamefont {K.}~\bibnamefont
  {Hammerer}}, \bibinfo {author} {\bibfnamefont {E.~M.}\ \bibnamefont {Rasel}},
  \ and\ \bibinfo {author} {\bibfnamefont {N.}~\bibnamefont {Gaaloul}},\ }\href
  {\doibase 10.1140/epjd/s10053-021-00069-9} {\bibfield  {journal} {\bibinfo
  {journal} {The European Physical Journal D}\ }\textbf {\bibinfo {volume}
  {75}} (\bibinfo {year} {2021}),\ 10.1140/epjd/s10053-021-00069-9}\BibitemShut
  {NoStop}%
\bibitem [{\citenamefont {Schlippert}\ \emph {et~al.}(2021)\citenamefont
  {Schlippert}, \citenamefont {Meiners}, \citenamefont {Rengelink},
  \citenamefont {Schubert}, \citenamefont {Tell}, \citenamefont {Wodey},
  \citenamefont {Zipfel}, \citenamefont {Ertmer},\ and\ \citenamefont
  {Rasel}}]{Schlippert2020}%
  \BibitemOpen
  \bibfield  {author} {\bibinfo {author} {\bibfnamefont {D.}~\bibnamefont
  {Schlippert}}, \bibinfo {author} {\bibfnamefont {C.}~\bibnamefont {Meiners}},
  \bibinfo {author} {\bibfnamefont {R.}~\bibnamefont {Rengelink}}, \bibinfo
  {author} {\bibfnamefont {C.}~\bibnamefont {Schubert}}, \bibinfo {author}
  {\bibfnamefont {D.}~\bibnamefont {Tell}}, \bibinfo {author} {\bibfnamefont
  {{\'{E}}.}~\bibnamefont {Wodey}}, \bibinfo {author} {\bibfnamefont
  {K.}~\bibnamefont {Zipfel}}, \bibinfo {author} {\bibfnamefont
  {W.}~\bibnamefont {Ertmer}}, \ and\ \bibinfo {author} {\bibfnamefont
  {E.}~\bibnamefont {Rasel}},\ }in\ \href {\doibase 10.1142/9789811213984_0010}
  {\emph {\bibinfo {booktitle} {{CPT} and Lorentz Symmetry}}}\ (\bibinfo
  {publisher} {{WORLD} {SCIENTIFIC}},\ \bibinfo {year} {2021})\BibitemShut
  {NoStop}%
\bibitem [{\citenamefont {Corgier}\ \emph
  {et~al.}(2021{\natexlab{a}})\citenamefont {Corgier}, \citenamefont
  {Pezz\`e},\ and\ \citenamefont {Smerzi}}]{Corgier21PRA}%
  \BibitemOpen
  \bibfield  {author} {\bibinfo {author} {\bibfnamefont {R.}~\bibnamefont
  {Corgier}}, \bibinfo {author} {\bibfnamefont {L.}~\bibnamefont {Pezz\`e}}, \
  and\ \bibinfo {author} {\bibfnamefont {A.}~\bibnamefont {Smerzi}},\ }\href
  {\doibase 10.1103/PhysRevA.103.L061301} {\bibfield  {journal} {\bibinfo
  {journal} {Phys. Rev. A}\ }\textbf {\bibinfo {volume} {103}},\ \bibinfo
  {pages} {L061301} (\bibinfo {year} {2021}{\natexlab{a}})}\BibitemShut
  {NoStop}%
\bibitem [{\citenamefont {Corgier}\ \emph
  {et~al.}(2021{\natexlab{b}})\citenamefont {Corgier}, \citenamefont {Gaaloul},
  \citenamefont {Smerzi},\ and\ \citenamefont {Pezz\`e}}]{Corgier21PRL}%
  \BibitemOpen
  \bibfield  {author} {\bibinfo {author} {\bibfnamefont {R.}~\bibnamefont
  {Corgier}}, \bibinfo {author} {\bibfnamefont {N.}~\bibnamefont {Gaaloul}},
  \bibinfo {author} {\bibfnamefont {A.}~\bibnamefont {Smerzi}}, \ and\ \bibinfo
  {author} {\bibfnamefont {L.}~\bibnamefont {Pezz\`e}},\ }\href {\doibase
  10.1103/PhysRevLett.127.183401} {\bibfield  {journal} {\bibinfo  {journal}
  {Phys. Rev. Lett.}\ }\textbf {\bibinfo {volume} {127}},\ \bibinfo {pages}
  {183401} (\bibinfo {year} {2021}{\natexlab{b}})}\BibitemShut {NoStop}%
\bibitem [{\citenamefont {Anders}\ \emph {et~al.}(2021)\citenamefont {Anders},
  \citenamefont {Idel}, \citenamefont {Feldmann}, \citenamefont {Bondarenko},
  \citenamefont {Loriani}, \citenamefont {Lange}, \citenamefont {Peise},
  \citenamefont {Gersemann}, \citenamefont {Meyer-Hoppe}, \citenamefont
  {Abend}, \citenamefont {Gaaloul}, \citenamefont {Schubert}, \citenamefont
  {Schlippert}, \citenamefont {Santos}, \citenamefont {Rasel},\ and\
  \citenamefont {Klempt}}]{Anders21PRL}%
  \BibitemOpen
  \bibfield  {author} {\bibinfo {author} {\bibfnamefont {F.}~\bibnamefont
  {Anders}}, \bibinfo {author} {\bibfnamefont {A.}~\bibnamefont {Idel}},
  \bibinfo {author} {\bibfnamefont {P.}~\bibnamefont {Feldmann}}, \bibinfo
  {author} {\bibfnamefont {D.}~\bibnamefont {Bondarenko}}, \bibinfo {author}
  {\bibfnamefont {S.}~\bibnamefont {Loriani}}, \bibinfo {author} {\bibfnamefont
  {K.}~\bibnamefont {Lange}}, \bibinfo {author} {\bibfnamefont
  {J.}~\bibnamefont {Peise}}, \bibinfo {author} {\bibfnamefont
  {M.}~\bibnamefont {Gersemann}}, \bibinfo {author} {\bibfnamefont
  {B.}~\bibnamefont {Meyer-Hoppe}}, \bibinfo {author} {\bibfnamefont
  {S.}~\bibnamefont {Abend}}, \bibinfo {author} {\bibfnamefont
  {N.}~\bibnamefont {Gaaloul}}, \bibinfo {author} {\bibfnamefont
  {C.}~\bibnamefont {Schubert}}, \bibinfo {author} {\bibfnamefont
  {D.}~\bibnamefont {Schlippert}}, \bibinfo {author} {\bibfnamefont
  {L.}~\bibnamefont {Santos}}, \bibinfo {author} {\bibfnamefont
  {E.}~\bibnamefont {Rasel}}, \ and\ \bibinfo {author} {\bibfnamefont
  {C.}~\bibnamefont {Klempt}},\ }\href {\doibase
  10.1103/PhysRevLett.127.140402} {\bibfield  {journal} {\bibinfo  {journal}
  {Phys. Rev. Lett.}\ }\textbf {\bibinfo {volume} {127}},\ \bibinfo {pages}
  {140402} (\bibinfo {year} {2021})}\BibitemShut {NoStop}%
\bibitem [{\citenamefont {Kovachy}\ \emph {et~al.}(2015)\citenamefont
  {Kovachy}, \citenamefont {Hogan}, \citenamefont {Sugarbaker}, \citenamefont
  {Dickerson}, \citenamefont {Donnelly}, \citenamefont {Overstreet},\ and\
  \citenamefont {Kasevich}}]{Kovachy15PRL}%
  \BibitemOpen
  \bibfield  {author} {\bibinfo {author} {\bibfnamefont {T.}~\bibnamefont
  {Kovachy}}, \bibinfo {author} {\bibfnamefont {J.~M.}\ \bibnamefont {Hogan}},
  \bibinfo {author} {\bibfnamefont {A.}~\bibnamefont {Sugarbaker}}, \bibinfo
  {author} {\bibfnamefont {S.~M.}\ \bibnamefont {Dickerson}}, \bibinfo {author}
  {\bibfnamefont {C.~A.}\ \bibnamefont {Donnelly}}, \bibinfo {author}
  {\bibfnamefont {C.}~\bibnamefont {Overstreet}}, \ and\ \bibinfo {author}
  {\bibfnamefont {M.~A.}\ \bibnamefont {Kasevich}},\ }\href {\doibase
  10.1103/physrevlett.114.143004} {\bibfield  {journal} {\bibinfo  {journal}
  {Physical Review Letters}\ }\textbf {\bibinfo {volume} {114}} (\bibinfo
  {year} {2015}),\ 10.1103/physrevlett.114.143004}\BibitemShut {NoStop}%
\bibitem [{\citenamefont {Fattori}\ \emph {et~al.}(2008)\citenamefont
  {Fattori}, \citenamefont {D'Errico}, \citenamefont {Roati}, \citenamefont
  {Zaccanti}, \citenamefont {Jona-Lasinio}, \citenamefont {Modugno},
  \citenamefont {Inguscio},\ and\ \citenamefont {Modugno}}]{Fattori08PRL}%
  \BibitemOpen
  \bibfield  {author} {\bibinfo {author} {\bibfnamefont {M.}~\bibnamefont
  {Fattori}}, \bibinfo {author} {\bibfnamefont {C.}~\bibnamefont {D'Errico}},
  \bibinfo {author} {\bibfnamefont {G.}~\bibnamefont {Roati}}, \bibinfo
  {author} {\bibfnamefont {M.}~\bibnamefont {Zaccanti}}, \bibinfo {author}
  {\bibfnamefont {M.}~\bibnamefont {Jona-Lasinio}}, \bibinfo {author}
  {\bibfnamefont {M.}~\bibnamefont {Modugno}}, \bibinfo {author} {\bibfnamefont
  {M.}~\bibnamefont {Inguscio}}, \ and\ \bibinfo {author} {\bibfnamefont
  {G.}~\bibnamefont {Modugno}},\ }\href {\doibase
  10.1103/PhysRevLett.100.080405} {\bibfield  {journal} {\bibinfo  {journal}
  {Phys. Rev. Lett.}\ }\textbf {\bibinfo {volume} {100}},\ \bibinfo {pages}
  {080405} (\bibinfo {year} {2008})}\BibitemShut {NoStop}%
\bibitem [{\citenamefont {Kim}\ \emph {et~al.}(2022)\citenamefont {Kim},
  \citenamefont {Krzyzanowska}, \citenamefont {Henderson}, \citenamefont {Ryu},
  \citenamefont {Timmermans},\ and\ \citenamefont {Boshier}}]{Kim22arXiv}%
  \BibitemOpen
  \bibfield  {author} {\bibinfo {author} {\bibfnamefont {H.}~\bibnamefont
  {Kim}}, \bibinfo {author} {\bibfnamefont {K.}~\bibnamefont {Krzyzanowska}},
  \bibinfo {author} {\bibfnamefont {K.~C.}\ \bibnamefont {Henderson}}, \bibinfo
  {author} {\bibfnamefont {C.}~\bibnamefont {Ryu}}, \bibinfo {author}
  {\bibfnamefont {E.}~\bibnamefont {Timmermans}}, \ and\ \bibinfo {author}
  {\bibfnamefont {M.}~\bibnamefont {Boshier}},\ }\href@noop {} {\enquote
  {\bibinfo {title} {One second interrogation time in a 200 round-trip
  waveguide atom interferometer},}\ } (\bibinfo {year} {2022}),\ \Eprint
  {http://arxiv.org/abs/2201.11888} {arXiv:2201.11888 [physics.atom-ph]}
  \BibitemShut {NoStop}%
\end{thebibliography}%
\end{document}